\documentclass[12pt]{article}
\pdfoutput=1

\usepackage{draft} 
\usepackage{hyperref}
\usepackage{graphicx,color,subfig}
\usepackage{cite}
\usepackage{mciteplus}
\usepackage{skak}
\usepackage{empheq}
\DeclareFontFamily{OT1}{pzc}{}
\DeclareFontShape{OT1}{pzc}{m}{it}{<-> s * [1.10] pzcmi7t}{}
\DeclareMathAlphabet{\mathpzc}{OT1}{pzc}{m}{it}

\def\be#1\ee{\begin{align}#1\end{align}}

\begin{document}

\unitlength = .8mm

\begin{titlepage}

\begin{center}

\hfill \\
\hfill \\
\vskip 1cm

\title{Recursive Representations of Arbitrary Virasoro Conformal Blocks}

\author{Minjae Cho, Scott Collier, Xi Yin}

\address{
Jefferson Physical Laboratory, Harvard University, \\
Cambridge, MA 02138 USA
}

\email{minjaecho@fas.harvard.edu, scollier@physics.harvard.edu, xiyin@fas.harvard.edu}

\end{center}

\abstract{
We derive recursive representations in the internal weights of $N$-point Virasoro conformal blocks in the sphere linear channel and the torus necklace channel, and recursive representations in the central charge of arbitrary Virasoro conformal blocks on the sphere, the torus, and higher genus Riemann surfaces in the plumbing frame.
}

\vfill

\end{titlepage}

\eject

\begingroup
\hypersetup{linkcolor=black}
\tableofcontents
\endgroup

\section{Introduction} 

A two-dimensional conformal field theory is characterized by its spectrum of Virasoro primaries and their OPE coefficients. Given these data, all correlation functions of the CFT on any Riemann surface can be constructed, through the Virasoro conformal blocks \cite{Belavin:1984vu, Moore:1988qv} which sum up all descendant contributions of the conformal families in consideration. Direct evaluation of the conformal blocks based on the definition by summing over Virasoro descendants is computationally costly and is practically intractable beyond the first few levels even with computer algebra. 

An efficient method for computing the sphere 4-point Virasoro conformal block was found by Zamolodchikov in \cite{Zamolodchikov:1985ie}, in the form of a recurrence relation in the central charge $c$. This is based on the observation that the conformal block can be analytically continued as a meromorphic function in $c$, whose poles are dictated by degenerate representations of the Virasoro algebra, together with a simplification in the large $c$ limit where the Virasoro block reduces to that of the global conformal group $SL(2)$. An analogous recurrence formula through the analytic continuation in the internal weight $h$ rather than the central charge was subsequently found in \cite{1987TMP....73.1088Z}. These recurrence formulae have played an essential role both in computing string amplitudes \cite{Chang:2014jta, Lin:2015zea} and in the numerical conformal bootstrap approach to 2D CFTs \cite{Lin:2015wcg, Lin:2016gcl, Collier:2017shs} (in \cite{Collier:2017shs}, for instance, the explicit expansion of a Virasoro conformal block to level 200 was used).

The recursive representations have also been extended to super-Virasoro conformal blocks \cite{Hadasz:2006qb, Hadasz:2008dt, Hadasz:2012im}, and to torus 1-point conformal blocks \cite{Hadasz:2009db,Alkalaev:2016fok}. More general Virasoro conformal blocks (higher points, higher genera) are important to the computation of certain string amplitudes as well as for more sophisticated numerical conformal bootstrap analyses. Our aim is to provide a complete set of recurrence relations for efficient evaluation of Virasoro conformal blocks on a Riemann surface of any genus with any number of external primary vertex operator insertions.

The main results of this paper are:

\noindent (1) {\it We extend the $c$-recursion relation to sphere and torus $N$-point Virasoro conformal blocks in all channels.}

The first key observation, which is common to all recurrence relations discussed in this paper, is that when we analytically continue in the central charge $c$ or the internal weights $h_i$, there is a pole whenever an internal Virasoro representation becomes degenerate and develops a null state at some level $rs$ \cite{Belavin:1984vu,DiFrancesco:639405}. The residue of this pole is proportional to the conformal block evaluated at the degenerate internal weight shifted by $rs$, with a universal coefficient that is a known function of the internal and external weights.

With this understanding, the determination of the recurrence relation boils down to identifying the large $c$ or large internal weight limits. The large $c$ limit of the sphere $N$-point Virasoro conformal block reduces to that of the global $SL(2)$ block, which is relatively easy to compute. The large $c$ limit of the torus $N$-point Virasoro conformal block turns out to reduce to the product of the torus vacuum character and a corresponding global $SL(2)$ block. The factorization property of the large central charge limit of the `light' block (with all weights held fixed) was originally observed in the case of the torus 1-point block in \cite{Alkalaev:2016fok}.
\begin{figure}[h!]
\center{\subfloat{\includegraphics[width=.42\textwidth]{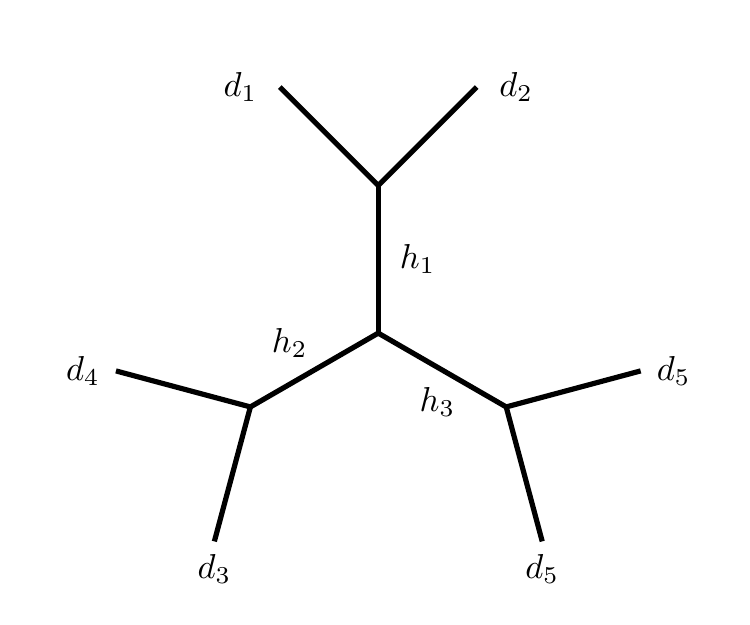}}~~
\subfloat{\includegraphics[width=.42\textwidth]{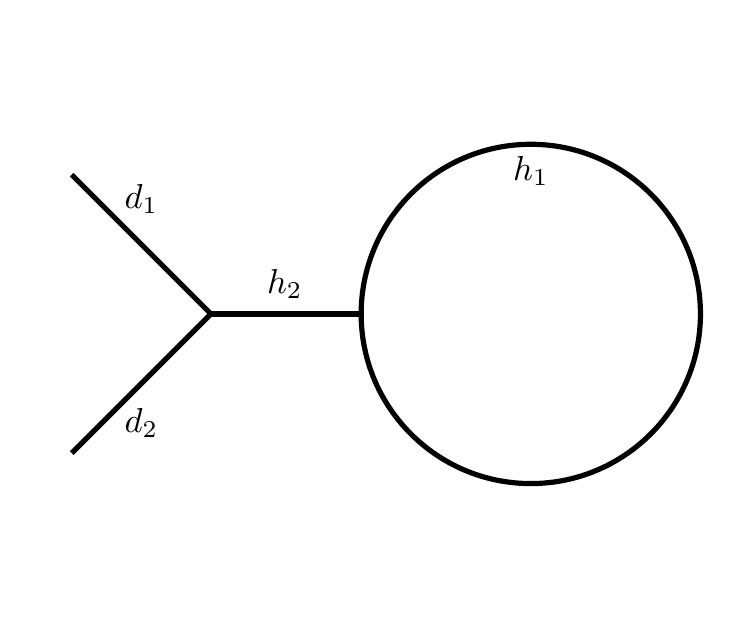}}}
\caption{The sphere six-point block in the trifundamental channel (left) and the torus two-point block in the OPE channel (right). Our $c$-recursion representation for arbitrary sphere and torus $N$-point blocks enables recursive evaluation of these blocks; we work these cases out explicitly in Section \ref{sec:cRecursionExamples}.}\label{fig:BlockExamples}
\end{figure}

\bigskip

\noindent (2) {\it We find the $h$-recursion for the sphere $N$-point Virasoro blocks in the linear channel, and torus $N$-point Virasoro blocks in the necklace channel.}

\begin{figure}[h!]
\subfloat{\includegraphics[width=.32\textwidth]{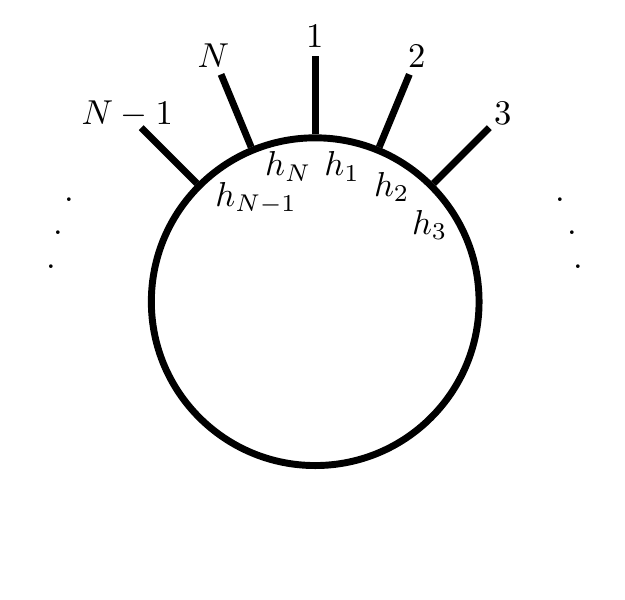}}~~
\subfloat{\includegraphics[width=.6\textwidth]{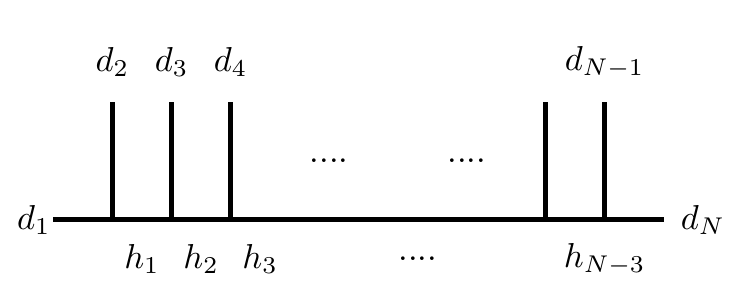}}
\caption{The torus $N$-point block in the necklace channel (left) and the sphere $N$-point block in the linear channel (right).}\label{fig:necklacelinear}
\end{figure}

To determine the $h$-recursion relations, we need to know the behavior of the Virasoro conformal block in suitable large internal weight limits, which turns out to be very subtle. In the case of the torus $N$-point block in the necklace channel, for instance, the simplification occurs when all internal weights $h_i$ along the necklace are taken to infinity simultaneously, with $h_i-h_j$ kept finite. In this limit, the necklace conformal block reduces to a non-degenerate torus Virasoro character. This observation is powerful enough to determine the recurrence relation for the necklace conformal block.

A degeneration limit of the torus $(N-1)$-point necklace block gives the sphere $N$-point conformal block in the linear channel. In the latter case, our recurrence relation makes use of the limit where all the internal weights $h_i$ and a pair of external weights $d_1$ and $d_N$ are taken to infinity simultaneously along a line that begins on $d_1$ and ends on $d_N$, again with their differences $h_i-h_j, h_i-d_1$, and $h_i-d_N$ kept finite. Note that this is {\it different from} Zamolodchikov's $h$-recurrence relation in the sphere 4-point case, where the recursion only applies to the internal weight. In particular, in our formulation of the sphere $N$-point recursion in the linear channel, it suffices to work with the standard cross ratios rather than Zamolodchikov's elliptic nome.

\bigskip

\noindent (3) {\it We give a complete set of recipes for the $c$-recursion relation for the most general $N$-point Virasoro conformal blocks on a genus $g$ Riemann surface, based on a plumbing construction through a given pair-of-pants decomposition.}

In formulating the higher genus Virasoro conformal blocks, based on a particular pair-of-pants decomposition, one must choose a conformal frame defined by a choice of the fundamental domain and gluing maps along its boundaries. Differences in the choice of conformal frame not only lead to different parameterizations of the moduli, but also extra factors multiplying the conformal block due to the conformal anomaly. We choose to construct the (punctured) Riemann surface by gluing together 3-holed Riemann spheres, represented by 2-holed discs on the complex plane, with $SL(2,\mathbb{C})$ M\"obius maps along their boundary components. Formally, since only $SL(2)$ maps are used in such a plumbing construction, it also makes sense to define a corresponding global $SL(2)$ block, by summing up $L_{-1}$ descendants at the holes.

We will show that in this frame, the genus $g$, $N$-point Virasoro conformal block remains finite in the $c\to \infty$ limit. In particular, the same is true for the genus $g$ vacuum block, whose large $c$ limit is expected to exponentiate into the form $e^{-c {\cal F}_0}$ to leading order, where ${\cal F}_0$ is the holomorphic part of a suitably regularized Einstein-Hilbert action on a hyperbolic handlebody \cite{Krasnov:2000zq, Yin:2007gv}. In our frame, ${\cal F}_0$ is simply zero, and the $c\to \infty$ limit of the vacuum block is finite. Further, the finite part of the $c\to \infty$ vacuum block is given by the 1-loop partition function of 3D pure gravity on the hyperbolic handlebody, as computed in \cite{Giombi:2008vd}.

We will show that the $c\to \infty$ limit of the genus $g$ Virasoro conformal block factorizes into the product of the $c\to \infty$ vacuum block and the global $SL(2)$ block defined through the above mentioned plumbing construction. This is a generalization of the factorization property of the light block at large central charge first proven in the case of the torus 1-point block in \cite{Alkalaev:2016fok}.
\begin{figure}[h!]
\includegraphics[width=.99\textwidth]{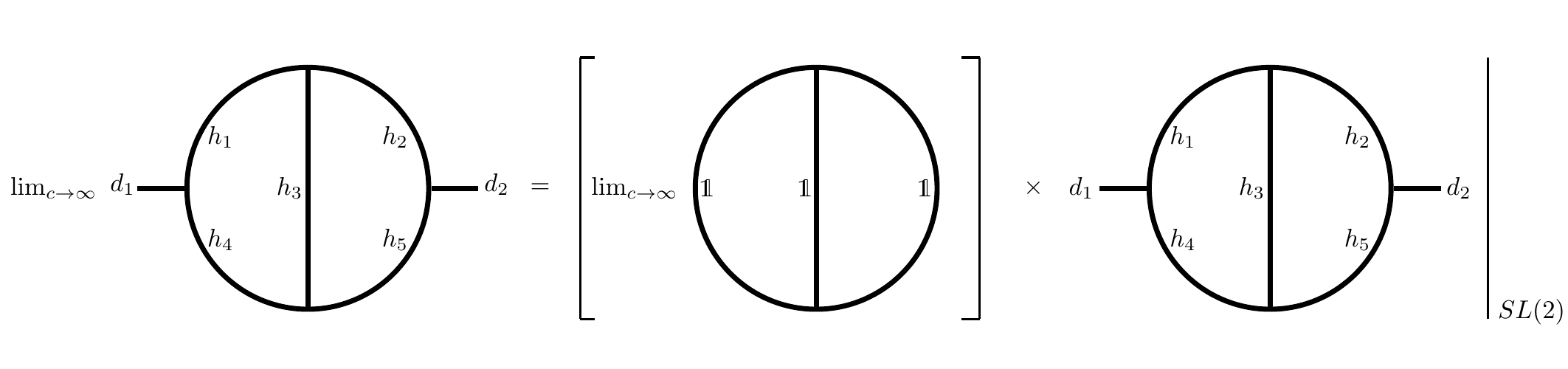}
\caption{The large-$c$ factorization of a genus-2 two-point block in the plumbing frame, in which the punctured Riemann surface is built by plumbing together two-holed (or punctured) discs using $SL(2)$ maps.}\label{fig:factorization}
\end{figure}

\bigskip

The paper is organized as follows. In section \ref{S4pt} we review Zamolodchikov's recurrence relations for the sphere 4-point Virasoro conformal block. The $h$-recurrence relations for torus $N$-point necklace channel conformal block and the sphere $N$-point linear channel conformal block are derived in section \ref{htorus}. In section \ref{ctorus}, we formulate and prove the $c$-recurrence relation for sphere and torus $N$-point Virasoro conformal blocks in arbitrary channels. The generalization to higher genus is presented in section \ref{highergenussection}. We conclude in section \ref{discussion} with a discussion of potential applications of our results, and issues concerning the mapping of moduli parameters for the higher genus conformal blocks.

\section{Review of the sphere 4-point Virasoro block}\label{S4pt}

In this section, we review the recursive representations of the sphere 4-point Virasoro conformal blocks, originally derived in \cite{Zamolodchikov:1985ie, 1987TMP....73.1088Z}. We follow the notations and derivations of \cite{Teschner:2001rv,Hadasz:2006qb, Hadasz:2009db, SuchanekThesis} in detail, as we will generalize their features to higher-point cases in later sections.

\subsection{Definition of Virasoro conformal block}
\label{defvirasoro}

Using the global $SL(2,\bC)$ invariance, the 4-point function of Virasoro primaries of weight $(d_i,\bar{d}_i)$, $i=1,...,4$, on the Riemann sphere can be brought to the form
\ie
\langle \phi_4'(\infty,\infty)\phi_3(1,1)\phi_2(z,\bar{z})\phi_1(0,0)\rangle=\langle\n_4\otimes\bar{\n}_4|\phi_3(1,1)\phi_2(z,\bar{z})|\n_1\otimes\bar{\n}_1\rangle,
\fe
where $\phi'(\infty,\infty) = \lim_{w,\bar w\to\infty}w^{2d_4}\bar w^{2\bar d_4}\phi_4(w,\bar w)$, $|\n_i\otimes\bar{\n}_i\rangle$ is the state corresponding to the primary operator $\phi_i$ inserted at the origin in radial quantization, and $\langle \n_i\otimes\bar{\n}_i|$ is the BPZ conjugate. Inserting a complete set of states in between $\phi_2$ and $\phi_3$, we can write
\ie
&\langle\n_4\otimes\bar{\n}_4|\phi_3(1,1)\phi_2(z,\bar{z})|\n_1\otimes\bar{\n}_1\rangle=
\\
&\sum_{h,\bar{h}}\sum_{\substack{|M|=|N|=n\\|P|=|Q|=m}}\langle\n_4\otimes\bar{\n}_4|\phi_3(1,1)|L_{-N}\n_{h}\otimes L_{-P}\bar{\n}_{\bar{h}}\rangle(G^n_{c,h})^{NM}(G^m_{c,\bar{h}})^{PQ}\langle L_{-M}\n_{h}\otimes L_{-Q}\bar{\n}_{\bar{h}}|\phi_2(z,\bar{z})|\n_1\otimes\bar{\n}_1\rangle.
\fe
Let us explain the notations here. The first sum is over the spectrum of Virasoro primaries of weights\footnote{To avoid overly cluttered notation, we have omitted the labels of possibly degenerate primaries, which can be restored easily when necessary.} $(h,\bar{h})$ and the second sum is over descendants in the corresponding conformal family. $M,N,P,Q$ are integer partitions in descending order that label Virasoro descendants. $L_{-N}$ stands for a chain of Virasoro generators corresponding to a specific partition $N$ of the non-negative integer $n=|N|$. For example, $N=\{2,1,1\}$ with $|N|=4$ gives rise to $L_{-N}=L_{-2}L_{-1}L_{-1}$. $G^n_{c,h}$ is the Gram matrix at level $n$ for a weight $h$ representation of the Virasoro algebra of central charge $c$, and $\left(G^n_{c,h}\right)^{NM}$ stands for the inverse Gram matrix element.

We will make extensive use of the 3-point function of general Virasoro descendants, which factorizes into its holomorphic and anti-holomorhic parts, of the form \cite{Teschner:2001rv}
\ie\label{3ptnotation}
\langle\x_3\otimes\bar{\x}_3|V_2(z,\bar{z})|\x_1\otimes\bar{\x}_1\rangle=C_{321}\R(\x_3,\x_2,\x_1|z)\R(\bar{\x}_3,\bar{\x}_2,\bar{\x}_1|\bar{z}).
\fe
Here $V_i$ represents a general Virasoro descendant of the primary $\phi_i$, while $(\x_i, \bar\x_i)$ label the corresponding states in the Verma modules associated with the left and right Virasoro algebra. The structure constant $C_{321}$ is the coefficient of the 3-point function of the primaries $\phi_3, \phi_2, \phi_1$. $\R(\x_3,\x_2,\x_1|z)$ is determined entirely by the Virasoro algebra in terms of the weights of the primaries, as we briefly review in Appendix \ref{app:review}; in particular, for primary states $\nu_i$, we have $\R(\n_3,\n_2,\n_1|z=1)=1$.
Using this notation, the 4-point function can be written as
\ie
\langle\n_4\otimes\bar{\n}_4|\phi_3(1,1)\phi_2(z,\bar{z})|\n_1\otimes\bar{\n}_1\rangle=\sum_{h,\bar{h}}C_{43(h,\bar{h})}C_{(h,\bar{h})21}&z^{h-d_2-d_1}F(h,z,d_1,d_2,d_3,d_4,c)
\\
\times&{\bar z}^{{\bar h}-{\bar d}_2-{\bar d}_1}F({\bar h},{\bar z},{\bar d}_1,{\bar d}_2,{\bar d}_3,{\bar d}_4,c),
\fe
where $F(h,z,d_1,d_2,d_3,d_4,c)$ is the holomorphic Virasoro conformal block of interest,
\ie\label{4ptBlock}
F(h,z,d_1,d_2,d_3,d_4,c)=\sum_{|N|=|M|=n\geq 0} z^n\R(\n_4,\n_3,L_{-N}\n_h|1)\left(G^n_{c,h}\right)^{NM}\R(L_{-M}\n_h,\n_2,\n_1|1).
\fe

Let us note a subtlety in our convention of $\rho(\xi_3, \xi_2, \xi_1|z)$ that will become particularly important later for the torus and higher genus conformal blocks. In the definition of this 3-point function, $\xi_1$ and $\xi_2$ are Virasoro descendants of the form $L_{-N_1} |h_1\rangle$ and $L_{-N_2}|h_2\rangle$ inserted on the complex plane at $0$ and $z$, while $\xi_3$ is the BPZ conjugate of a state of the form $L_{-N_3}|h_3\rangle$, inserted at $\infty$. In constructing a more general conformal block, associated with a pair-of-pants decomposition of a punctured Riemann surface, we will be contracting such 3-point functions of descendants with inverse Gram matrices. This corresponds to a plumbing construction where we cut out holes centered at 0, $z$, and $\infty$ on the complex plane, resulting in 2-holed discs, and identify boundary components of pairs of 2-holed discs via $SL(2,\mathbb{C})$ M\"obius maps. This amounts to a choice of conformal frame for the conformal block in question, which turns out to be particularly convenient for the $c$-recursive representation to be discussed later. A different choice of frame would generally lead to a conformal block that differs by a factor of the conformal anomaly.

One could consider a different 3-point function of descendants, $\tilde\rho(\xi_3, \xi_2, \xi_1|w)$, defined as the matrix element of the Virasoro descendant $\xi_2$ inserted at position $w$ on the cylinder $w\sim w+2\pi$, between the states $\langle \xi_3|$ and $|\xi_1\rangle$ on the cylinder (say both defined at ${\rm Im} w=0$). While the cylinder can be conformally mapped to the complex plane via $z=e^{-iw}$, $\xi_2$ being a descendant does not transform covariantly. For instance, $\tilde\rho(\xi_3, \xi_2, \xi_1|w=0)$ coincides with $\rho(\xi_3, \xi_2, \xi_1|z=1)$ when $\xi_2$ is a primary, but not otherwise. For certain conformal blocks it may be convenient to use a plumbing construction based on gluing together 1-holed cylinders rather than 2-holed discs, which would amount to contracting 3-point functions like $\tilde\rho(\xi_3, \xi_2, \xi_1|0)$ rather than $\rho(\xi_3, \xi_2, \xi_1|1)$ with inverse Gram matrices. This would result in the block in a different conformal frame.

\subsection{Simple pole structure and its residue}\label{Res}

Let us now consider the analytic continuation of the Virasoro conformal block in $h$ and in $c$. The presence of the inverse Gram matrix in (\ref{4ptBlock}) introduces simple poles, corresponding to the values of $h$ and $c$ where the Virasoro representation admits a null state at the corresponding level. Therefore, one can write
\ie\label{pole}
F(h,z,d_1,d_2,d_3,d_4,c)&=f_h(h,z,d_1,d_2,d_3,d_4,c)+\sum_{r\geq1,s\geq1}{V_{rs}(z,d_1,d_2,d_3,d_4,c)\over h-d_{rs}(c)}
\\
&=f_c(h,z,d_1,d_2,d_3,d_4,c)+\sum_{r\geq2,s\geq1}{W_{rs}(z,h,d_1,d_2,d_3,d_4)\over c-c_{rs}(h)},
\fe
where $f_h$ and $f_c$ are entire holomorphic functions in $h$ and in $c$ respectively. In the first line, we have assumed a generic value of $c$, whereas in the second we have assumed a generic value of $h$. The pole positions $d_{rs}(c)$ and $c_{rs}(h)$ are \cite{Belavin:1984vu,DiFrancesco:639405}\footnote{As a subscript, $rs$ stands for separate labels $r$ and $s$, not to be confused with the product $rs$.}
\ie\label{eq:MinimalWeights}
&d_{rs}(c)={(b+b^{-1})^2\over4}-{(rb+sb^{-1})^2\over4}~~\text{with}~c=1+6(b+b^{-1})^2,~~~r=1,2,3,...,~s=1,2,3,...
\\
&c_{rs}(h)=1+6(b_{rs}(h)+b_{rs}(h)^{-1})^2~~\text{with}~b_{rs}(h)^2={rs-1+2h+\sqrt{(r-s)^2+4(rs-1)h+4h^2}\over 1-r^2},
\\
&~~~~~~~~~~~~~~~~~~~~~~~~~~~~~~~~~~~~~~~~~~~~~~~~~~~~~~~~~~~~~~~~~~~~~~~~~~~~~~~~~~~ r=2,3,4,...,~s=1,2,3,...
\fe
Note that the two types of residues $V_{rs}$ and $W_{rs}$ are related by
\ie\label{hcresidue}
W_{rs}(z,h,d_i)=-{\partial c_{rs}(h)\over\partial h}V_{rs}(z,d_i,c=c_{rs}(h)).
\fe

The Verma module of the degenerate primary of weight $d_{rs}$ contains a null descendant at level $rs$. In the degeneration limit $h\to d_{rs}$, a new primary emerges at level $rs$ in place of the null state, which generates a sub-Verma module. The key observation in \cite{Zamolodchikov:1985ie} was that the residue at $h=d_{rs}$ is proportional to the Virasoro block whose internal representation is given by this sub-Verma module, namely one with internal weight $d_{rs}+rs$. This can be seen from (\ref{4ptBlock}) as follows. Following \cite{Hadasz:2006qb, Hadasz:2009db, SuchanekThesis}, we write the null descendant at level $rs$ corresponding to $d_{rs}$ as
\ie\label{null}
\X_{rs}=\sum_{|M|=rs}\X_{rs}^ML_{-M}\n_{d_{rs}},
\fe
where the normalization convention is such that the coefficient $\chi_{rs}^{\{1,1,\cdots, 1\}}$ of $L_{-1}^{rs}$ is equal to 1. For any Verma module associated to a primary of weight $h$, one can choose a basis for the level $rs$ and higher descedants that includes the states
\ie\label{nullbasis}
L_{-N}\X_{rs}^h~~\text{with}~~\X_{rs}^h\equiv \sum_{M}\X_{rs}^ML_{-M}\n_h.
\fe
Here, $\X_{rs}^M$ is the coefficient that appears in (\ref{null}), whereas $\chi^h_{rs}$ denotes a state (at level $rs$, which is not null for generic $h$). Other basis states are chosen generically. By definition, $\text{lim}_{h\rightarrow d_{rs}}\X_{rs}^h = \X_{rs}$.  The residue $V_{rs}$ in (\ref{pole}) receives contributions only from descendants of the form $L_{-N}\X^h_{rs}$ (whose level is $rs+|N|$), and is given by
\ie\label{residue}
V_{rs}(z,d_i,c)=&\lim_{h\rightarrow d_{rs}}(h-d_{rs})F(h,z,d_i,c)
\\
=&A_{rs}^cz^{rs}\sum_{|N|=|M|=n\geq 0}z^n\R(\n_4,\n_3,L_{-N}\X_{rs}|1)\left(G^n_{c,d_{rs}+rs}\right)^{NM}\R(L_{-M}\X_{rs},\n_2,\n_1|1),
\fe
where
\ie\label{eq:Ars}
A_{rs}^c=\lim_{h\rightarrow d_{rs}}\left({\langle\X_{rs}^h|\X_{rs}^h\rangle\over h-d_{rs}}\right)^{-1}={1\over2}\prod_{m=1-r}^r\prod_{n=1-s}^s(mb+nb^{-1})^{-1}, ~~(m,n)\neq(0,0),(r,s)
\fe
is guessed in \cite{Zamolodchikov:1985ie} and checked in \cite{Zamolodchikov:2003yb}. A key property that will be used repeatedly later is the factorization \cite{Hadasz:2006qb, Hadasz:2009db, SuchanekThesis}
\ie\label{factorization}
\R(L_{-M}\X_{rs},\n_2,\n_1|1)=\R(L_{-M}\n_{d_{rs}+rs},\n_2,\n_1|1)\R(\X_{rs},\n_2,\n_1|1).
\fe
Here, $\n_{d_{rs}+rs}$ stands for a primary of weight $d_{rs}+rs$. The second factor on the RHS is the fusion polynomial
\ie\label{eq:FusionPolynomial}
\R(\X_{rs},\n_2,\n_1|1)=&P^{rs}_c\begin{bmatrix}d_1\\d_2\end{bmatrix}
\\
=&\prod_{p=1-r ~{\rm step}~2}^{r-1}\, \prod_{q=1-s~{\rm step}~2}^{s-1}{\lambda_1+\lambda_2+pb+qb^{-1}\over2}{\lambda_1-\lambda_2+pb+qb^{-1}\over2},
\fe
where the products are taken over $p+r=1~ {\rm mod}~2,$ $q+s=1~{\rm mod}~2$, and $\lambda_i$ are defined by $d_i={1\over4}(b+b^{-1})^2-{1\over4}\lambda_i^2$. By plugging (\ref{factorization}) into (\ref{residue}) and comparing with (\ref{4ptBlock}), we determine the residue
\ie\label{Residue}
V_{rs}(z,d_i,c)=z^{rs}A_{rs}^cP^{rs}_c\begin{bmatrix}d_1\\d_2\end{bmatrix}P^{rs}_c\begin{bmatrix}d_4\\d_3\end{bmatrix}F(h\rightarrow d_{rs}+rs,z,d_i,c).
\fe
Indeed, the residue is proportional to the Virasoro conformal block with internal weight evaluated at the null descendant value $d_{rs}+rs$. This sets a recursive representation of the Virasoro block, once the regular term $f_h$ or $f_c$ in (\ref{pole}) is known. In particular, the presence of the factor $z^{rs}$ in (\ref{Residue}) allows for the determination of the coefficient at any given order in the power series expansion in $z$ by finitely many iterations of (\ref{pole}).

\subsection{Determining the regular part}

First, let us determine the regular part $f_c(h,z,d_i,c)$ in (\ref{pole}) by studying the conformal block in the large-$c$ limit. The latter is computable by inspecting the definition (\ref{4ptBlock}). It follows from Ward identities that the 3-point function of the form $\R(\n_4,\n_3,L_{-N}\n_h|1)$ is independent of $c$, simply because there are no non-$L_{-1}$ Virasoro generators acting on $\nu_4$ and $\nu_3$.
Meanwhile, the inverse Gram matrix elements are suppressed in the large $c$ limit, except for one matrix element that corresponds to the inner product of a pair of $L_{-1}^n$ descendants,
\ie
\lim_{c\rightarrow\infty}\left(G^n_{c,h}\right)^{L_{-1}^n L_{-1}^n}={1\over n!(2h)_n},
\fe
where $(a)_n\equiv a(a+1)(a+2)...(a+n-1)$ is the Pochhammer symbol. This gives the only level $n$ term in (\ref{4ptBlock}) that survives at $c\to \infty$. Using $\R(L_{-1}^n\n_h,\n_2,\n_1)=(h+d_2-d_1)_n$, we obtain the result
\ie\label{c4pt}
f_c(h,z,d_i,c)=\sum_{n=0}^\infty z^n {(h+d_2-d_1)_n(h+d_3-d_4)_n\over n!(2h)_n}={}_2F_1(h+d_2-d_1,h+d_3-d_4,2h,z).
\fe
In particular, $f_c$ is independent of $c$. This feature will make a reappearance in other cases to be considered later. 
It is often asserted that ``the large-$c$ limit of the Virasoro block is the global $SL(2)$ block", referring to the fact that only the contributions of the $L_{-1}^n$ descendants survive in the large $c$ limit here. We will see later that this is not true for the large $c$ limit of torus and higher genus Virasoro conformal blocks, but suitable modifications of the statement do hold. 

Together with $W_{rs}$ acquired by (\ref{hcresidue}) and (\ref{Residue}), we have a complete $c$-recursive representation of the sphere 4-point Virasoro conformal block
\ie\label{crecursion}
F(h,z,d_i,c)=&{}_2F_1(h+d_2-d_1,h+d_3-d_4,2h,z)
\\
&+\sum_{r\geq2,s\geq1}-{\partial c_{rs}(h)\over\partial h}{z^{rs}A_{rs}^{c_{rs}}\over c-c_{rs}(h)}P^{rs}_{c_{rs}}\begin{bmatrix}d_1\\d_2\end{bmatrix}P^{rs}_{c_{rs}}\begin{bmatrix}d_4\\d_3\end{bmatrix}F(h\rightarrow h+rs,z,d_i,c\rightarrow c_{rs}).
\fe

The story for the $h$-regular part $f_h$ is more complicated. In \cite{1987TMP....73.1088Z}, Zamolodchikov considered a semiclassical limit of large $c$ with ratios $c/h, c/d_i$ kept finite, where the conformal block is expected to be the exponential of a ``classical block" of order $c$. Through the monodromy equation related to the classical block, the large-$h$ behavior was determined as a function of the elliptic nome $q$, related to the cross ratio $z$ by $q=\exp\left({i\pi {K'(z)\over K(z)}}\right)$, where $K(z)$ is the complete elliptic integral of the first kind. The final answer is 
\ie\label{hrecursion}
&z^{h-d_1-d_2}F(h,z,d_i,c)
\\
=&(16q(z))^{h-{(c-1)\over 24}}z^{{(c-1)\over24}-d_1-d_2}(1-z)^{{(c-1)\over24}-d_2-d_3}\theta_3(q(z))^{{(c-1)\over2}-4(d_1+d_2+d_3+d_4)}H(c,h,d_i,q(z)),
\fe
where $H(c,h,d_i,q)$ is determined recursively,
\ie
H(c,h,d_i,q)=1+\sum_{rs\geq1}{(16q)^{rs}A_{rs}^c\over h-d_{rs}}P^{rs}_c\begin{bmatrix}d_1\\d_2\end{bmatrix}P^{rs}_c\begin{bmatrix}d_4\\d_3\end{bmatrix}H(c,h\rightarrow d_{rs}+rs,d_i,q).
\fe
An alternative viewpoint on the $q$-expansion was provided in \cite{Maldacena:2015iua}. There, the 4-punctured sphere was mapped to the ``pillow" geometry $T^2/\bZ_2$ with four corners. There is an external vertex operator insertion at each corner. The $q$-expansion has the natural interpretation in terms of matrix elements of the propagator along the pillow, between states created by pairs of vertex operators at the corners. The $q$-expansion of the Virasoro conformal block converges uniformly on the unit $q$-disc $|q|<1$, which extends beyond the complex $z$-plane; for this reason, it is typically preferred in evaluations at high precision such as in numerical bootstrap, as well as for analytic continuation to Lorentzian signature. The solution to the recursion relations was studied in \cite{Perlmutter:2015iya}.

At the moment, it is unclear whether there is a useful analog of the $q$-expansion for more general Virasoro conformal blocks (higher points, higher genus). In the next section, we will instead work with an expansion in the cross ratio $z$ for the sphere $N$-point block in the linear channel, and derive a recursion relation that involves simultaneous shifts of the internal weights and a pair of external weights. In particular, we will not derive the analog of $f_h$ in (\ref{pole}), but rather a different kind of large-weight limit of the conformal block. The specialization of our $h$-recursion formula to the sphere 4-point block case differs from Zamolodchikov's $h$-recursion in several ways: we do not make use of the elliptic nome, the regular (non-polar) part is very simple, but the recursion involves shifting both internal and external weights.

\section{$h$-recursion for torus $N$-point Virasoro conformal blocks in the necklace channel (and sphere $N$-point blocks in the linear channel)}\label{htorus}

\begin{figure}[h!]
\centering
\subfloat{\includegraphics[width=.49\textwidth]{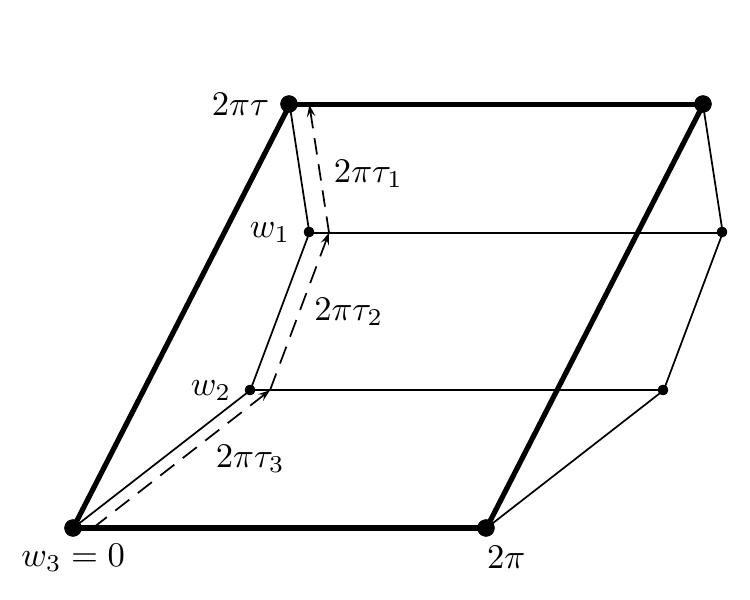}}
\caption{$N$-cylinder decomposition of necklace channel for $N=3$ case.}\label{fig:cylinderdecom}
\end{figure}

In this section, we derive recursion relations in the internal weights $h_i$ for torus $N$-point Virasoro blocks in the necklace channel for $N>1$. The $N=1$ case is studied in \cite{Hadasz:2009db} and we will discuss this case in the next section. The necklace channel is where complete sets of states are inserted in between every consecutive pair of external operators as shown at Figure~\ref{fig:necklacelinear}. 
This channel can also be viewed as the $N$-cylinder channel, where the torus is decomposed into $N$ cylinders, each of which contains exactly one external operator of weight $d_i$ at its origin. The $N=3$ case is illustrated in Figure~\ref{fig:cylinderdecom}. The result for the torus necklace channel reduces in a limit to the sphere block in the linear channel, thereby giving a recursion formula for the latter as well. 

\subsection{Definition of the Virasoro block in the necklace channel}

Consider a torus of modulus $\tau$, parameterized by a complex coordinate $z$, with the identification $z\sim z+2\pi \sim z+2\pi\tau$. We consider $N$ primary operators $\phi_i$ of weights $(d_i,\bar d_i)$ inserted at positions $z=w_i$ for $i=1,\cdots,N$. We set $w_N=0$ by convention, and write $w_i=2\pi(\T-\sum_{k=1}^i\T_k)$. In the necklace channel, the torus is decomposed into $N$ cylinders, of moduli $\tau_1, \tau_2, \cdots,\tau_N$, with $\sum_{k=1}^N\T_k=\T$. We will also write $q_i=e^{2\pi i\T_i}$.  The torus $N$-point function is decomposed in terms of Virasoro conformal blocks in this channel as
\ie
\langle O_1(w_1)O_2(w_2)...O_{N-1}(w_{N-1})O_N(0)\rangle_{T^2}=&\sum_{(h_1,{\bar h}_1),...,(h_N,{\bar h}_N)}\left(\prod_{i=1}^NC_{h_i,d_i,h_{i+1}}^{\bar{h}_i,\bar{d}_i,\bar{h}_{i+1}}q_i^{h_i-c/24}\bar{q}_i^{\bar{h}_i-c/24}\right)
\\
&
\times F\left(q_1,h_1,d_1,...,q_N,h_N,d_N,c\right)F\left(\bar{q}_1,\bar{h}_1,\bar{d}_1,...,\bar{q}_N,\bar{h}_N,\bar{d}_N,c\right).
\fe
Here $(h_i, \bar h_i)$ are the weights of the internal primaries. $F\left(q_1,h_1,d_1,...,q_N,h_N,d_N,c\right)$ is the holomorphic torus $N$-point necklace conformal block,
\ie\label{necklaceblock}
&F\left(q_1,h_1,d_1,...,q_N,h_N,d_N,c\right)
\\
=&\sum_{n_1,...,n_N=0}^{\infty}\left(\prod_{i=1}^{N}q_i^{n_i}\right)\sum_{|A_1|=|B_1|=n_1}...\sum_{|A_N|=|B_N|=n_N}\left[\prod_{i=1}^N\left(G^{n_i}_{h_i}\right)^{A_iB_i}\R(L_{-B_i}h_i,d_i,L_{-A_{i+1}}h_{i+1})\right].
\fe
The index $i$ ranges from 1 to $N$ cyclic, i.e. $i=N+1$ is identified with $i=1$. Here we have made use of an exponential mapping from each of the cylinders to the annulus, relating the matrix element of the primary $\phi_i$ at $w_i$ between a pair of descendant states to the 3-point function on the $z$-plane with the primary $\phi_i$ inserted at $z=1$. By a slight abuse of notation, in the sphere 3-point function of descendants $\rho$ we have labeled the primaries $\nu_i$ simply by their weights $h_i$, and have set $z=1$. 

\subsection{Polar part}\label{necklaceresidue}

Again due to the presence of the inverse Gram matrix, (\ref{necklaceblock}) has simple poles in $h_i$ or $c$ at values corresponding to degenerate Virasoro representations. Focusing on a single internal weight $h_i$, we have a simple pole expansion
\ie\label{ipole}
F=U_i+\sum_{1\leq r_is_i}{V_{r_is_i}\over h_i-d_{r_is_i}},
\fe
where $U_i$ is the $h_i$-regular part of the conformal block. The factor in (\ref{necklaceblock}) responsible for the pole at $h_i=d_{r_is_i}$ is
\ie
\R(L_{-B_{i-1}}h_{i-1},d_{i-1},L_{-A_i}h_i)\left(G^{n_i}_{h_i}\right)^{A_iB_i}\R(L_{-B_i}h_i,d_i,L_{-A_{i+1}}h_{i+1}).
\fe
In the limit $h_i\rightarrow d_{r_is_i}$, we can repeat the arguments in Section~\ref{Res}, now making use of a factorization property of the 3-point function involving null states that slightly generalizes (\ref{factorization})
\ie\label{factorization2}
\R(L_{-B}\X_{rs},\n_2,L_{-A}\n_1|1)=\R(L_{-B}\n_{d_{rs}+rs},\n_2,L_{-A}\n_1|1)\R(\X_{rs},\n_2,\n_1|1).
\fe
The derivation of this relation is discussed in Appendix \ref{app:3pt}. Therefore, the residue coefficient is captured by
\ie
&\R(L_{-B_{i-1}}h_{i-1},d_{i-1},L_{-A_i}\X_{r_is_i})\left(G^{n_i}_{d_{r_is_i}+r_is_i}\right)^{A_iB_i}\R(L_{-B_i}\X_{r_is_i},d_i,L_{-A_{i+1}}h_{i+1})
\\
=&\R(h_{i-1},d_{i-1},\X_{r_is_i})\R(\X_{r_is_i},d_i,h_{i+1})
\\
&\times\R(L_{-B_{i-1}}h_{i-1},d_{i-1},L_{-A_i}\n_{d_{r_is_i}+r_is_i})\left(G^{n_i}_{d_{r_is_i}+r_is_i}\right)^{A_iB_i}\R(L_{-B_i}\n_{d_{r_is_i}+r_is_i},d_i,L_{-A_{i+1}}h_{i+1}).
\fe
Following Section~\ref{Res}, and using the definition of the fusion polynomial, the residue in (\ref{ipole}) is determined to be
\ie\label{necklaceres}
V_{r_is_i}=q_i^{r_is_i}R_{r_is_i}(h_{i-1},h_{i+1},d_{i-1},d_i,c)F(h_i\rightarrow d_{r_is_i}+r_is_i),
\fe
with
\ie
R_{r_is_i}(h_{i-1},h_{i+1},d_{i-1},d_i,c)=A^c_{r_is_i}P_c^{r_is_i}\begin{bmatrix}h_{i-1}\\d_{i-1}\end{bmatrix}P_c^{r_is_i}\begin{bmatrix}h_{i+1}\\d_i\end{bmatrix}.
\fe

\subsection{Regular part}\label{hregular}

The $h_i$-regular part $U_i$ in (\ref{ipole}) is in fact quite complicated; fortunately, we do not need to compute $U_i$ directly. Let us define $a_i=h_i-h_1$, for $i=2,\cdots, N$, and consider the limit $h_1\rightarrow\infty$ with all $a_i$'s held fixed. In other words, we take the {\it simultaneous} large $h_i$ limit, with the differences $h_i-h_j$ kept finite. 
We will see that a drastic simplification of the conformal block occurs in this limit, giving rise to the regular part
\ie\label{largehnecklace}
F\left(q_1,h_1,d_1,...,q_N,h_N,d_N,c\right)\rightarrow \prod_{n=1}^{\infty}{1\over 1-(q_1q_2...q_N)^n},
\fe
which takes the form of a (non-degenerate) torus character. 

Let us begin with a basis of level $n$ descendants of a primary $|h\rangle$, of the form $L_{-A}|h\rangle$, where $A$ is a partition of the integer $n$ in descending order. We will write $|A|=n$, and $[A]$ for the number of Virasoro generators in $L_{-A}$ (the length of the partition). Note that in the large $h$, fixed $c$ limit, the inner product $\langle h| L_{-A}^\dagger L_{-B}|h\rangle$ scales like $h^{[A]}$ for $A=B$, no faster than $h^{[A]-1}$ for $[A]=[B]$, $A\not=B$, and no faster than $h^{{\rm min}([A],[B])}$ for $[A]\not=[B]$. We can thus construct via the Gram-Schmidt process an orthogonal basis of the form
\ie\label{orthob}
\ell_{-A} |h\rangle = L_{-A}|h\rangle + \sum_{|B|=n,~ [B]\leq [A],~B\not=A} f^A_B(c,h)L_{-B} |h\rangle,
\fe
such that 
\ie\label{fscaling}
& f^A_B(c,h) \sim {\cal O}(h^{-1}),~~~ [B]=[A],~B\not=A;
\\
& f^A_B(c,h) \sim {\cal O}(h^0),~~~ [B]<[A],
\fe
in the large $h$, fixed $c$ limit. The norm of the basis state $\ell_{-A}|h\rangle$ scales like
\ie
\langle h| \ell_{-A}^\dagger \ell_{-A} |h\rangle \sim h^{[A]}.
\fe

In the large $h_1$ limit with $a_i=h_i-h_1$ fixed ($i=2,\cdots,N$), the torus $N$-point block in the necklace channel (\ref{necklaceblock}) becomes
\ie\label{hlimit}
F\rightarrow\sum_{n_1,...,n_N=0}^{\infty}\left(\prod_{i=1}^{N}q_i^{n_i}\right)\sum_{|A_1|=n_1}...\sum_{|A_N|=n_N}\left[\prod_{i=1}^N{\R(\ell_{-A_i}h_1,d_i,\ell_{-A_{i+1}}h_1)\over\langle h_1|\ell_{-A_i}^\dag\ell_{-A_i}|h_1\rangle}\right].
\fe
Here we have traded every internal weight $h_i$ with $h_1$, which is valid to leading order. Let us investigate the large $h_1$ behavior of the numerator,
\ie
\R(\ell_{-A_i}h_1,d_i,\ell_{-A_{i+1}}h_1)=\sum_{|C|=|A_i|,~|B|=|A_{i+1}|}f^{A_i}_{C}f^{A_{i+1}}_{B}\R(L_{-C}h_1,d_i,L_{-B}h_1),
\fe
where we have extended the definition of $f^A_B$ in (\ref{orthob}) by setting $f^A_A=1$ (no summation over $A$) and $f^A_B=0$ for $[B]>[A]$. We can now evaluate the 3-point functions on the RHS using the Ward identities discussed in Appendix \ref{app:review}. Moving $L_{-C}$ to the right past $d_i$, one picks up commutator terms involving $\left[L_m,~\n_{d_i}\right]$, but the latter does not scale with $h_1$. Thus, to leading order in the large $h_1$ limit, we may freely move $L_{-C}$ through $d_i$ to obtain
\ie
\R(L_{-C}h_1,d_i,L_{-B}h_1)\sim\R(h_1,d_i,L_{-C}^\dagger L_{-B}h_1) \sim {\cal O}(h_1^{{\rm min}([B],[C])}).
\fe
It then follows from (\ref{fscaling}) that the terms in (\ref{hlimit}) that survive in the large $h_1$ limit have $A_1=A_2=\cdots=A_N$, with
\ie
{\R(\ell_{-A}h_1,d_i,\ell_{-A} h_1)\over\langle h_1|\ell_{-A}^\dag\ell_{-A}|h_1\rangle} \to 1.
\fe
Thus, the sum in (\ref{hlimit}) collapses to (\ref{largehnecklace}).

\subsection{$h$-recursion representation}

We can now combine the above results on the polar part and the large $h_1$, fixed $a_i$ asymptotics to obtain a complete recursive representation of torus $N$-point Virasoro conformal blocks in the necklace channel. First, we fix $a_i$ for $i=2,3,...,N$ and view the necklace block as a meromorphic function of $h_1$. Its simple pole expansion takes the form
\ie
F(q_1,h_1,d_1,q_2,a_2,d_1,...,q_N,a_N,d_N,c)= \prod_{n=1}^{\infty}{1\over 1-(q_1q_2...q_N)^n} +\sum_{i=1}^N\sum_{r_is_i\geq1}{B_{r_is_i}\over h_1+a_i-d_{r_is_i}},
\fe
where we have extended the definition of $a_i$ by including $a_1=0$. The residues $B_{r_is_i}$ are determined using (\ref{necklaceres}),
\ie
& B_{r_1s_1}=q_1^{r_1s_1}R_{r_1s_1}(d_{r_1s_1}+a_N,d_{r_1s_1}+a_2,d_N,d_1,c)
\\
&~~~~~~~~~~~~\times F(h_1\rightarrow d_{r_1s_1}+r_1s_1, a_i\rightarrow a_i-r_1s_1 ~\text{for}~ i=2,...,N),
\\
& B_{r_2s_2}=q_2^{r_2s_2}R_{r_2s_2}(d_{r_2s_2}-a_2,d_{r_2s_2}-a_2+a_3,d_1,d_2,c)F(h_1\rightarrow d_{r_2s_2}-a_2,a_2\rightarrow a_2+r_2s_2),
\\
& B_{r_Ns_N}=q_N^{r_Ns_N}R_{r_Ns_N}(d_{r_Ns_N}-a_N+a_{N-1},d_{r_Ns_N}-a_N,d_{N-1},d_N,c)
\\
& ~~~~~~~~~~~~\times F(h_1\rightarrow d_{r_Ns_N}-a_N,a_N\rightarrow a_N+r_Ns_N),
\\
& B_{r_is_i}=q_i^{r_is_i}R_{r_is_i}(d_{r_is_i}-a_i+a_{i-1},d_{r_is_i}-a_i+a_{i+1},d_{i-1},d_i,c)F(h_1\rightarrow d_{r_is_i}-a_i,a_i\rightarrow a_i+r_is_i)
\\
&~~~~~~~~~~~~~~~~~~~~~~~~~~~~~~~~~~~~~~~~~~~~~~~~~~~~~~~~~~~~~~~~~~~~~~~~~~~~~~~~~~~~~\text{for}~i=3,...,N-1.
\fe
We caution the reader that the shifted conformal blocks on the RHS still depend on the original $a_i=h_i-h_1$. While they are independent of $h_1$ as functions of $a_i$, they would still contain $h_1$ dependence when viewed as functions of the $h_i$'s.

Defining a reduced conformal block $f$ by factoring out the torus character,
\ie
F(q_1,h_1,d_1,q_2,a_2,d_1,...,q_N,a_N,d_N,c)= \left[ \prod_{n=1}^{\infty}{1\over 1-(q_1q_2...q_N)^n}\right] f(q_1,h_1,d_1,q_2,a_2,d_1,...,q_N,a_N,d_N,c),
\fe
we can express the recursion relation as
\ie\label{necklacefinal}
&f(q_1,h_1,d_1,q_2,a_2,d_1,...,q_N,a_N,d_N,c)
\\
=&1+\sum_{r_1s_1\geq1}{q_1^{r_1s_1}R_{r_1s_1}(d_{r_1s_1}+a_N,d_{r_1s_1}+a_2,d_N,d_1,c)\over h_1-d_{r_1s_1}}f(h_1\rightarrow d_{r_1s_1}+r_1s_1, a_i\rightarrow a_i-r_1s_1 ~\text{for}~ i=2,...,N)
\\
&+\sum_{i=2}^{N}\sum_{r_is_i\geq1}{q_i^{r_is_i}R_{r_is_i}(d_{r_is_i}-a_i+a_{i-1},d_{r_is_i}-a_i+a_{i+1},d_{i-1},d_i,c)\over h_1+a_i-d_{r_is_i}}f(h_1\rightarrow d_{r_is_i}-a_i,a_i\rightarrow a_i+r_is_i).
\fe
This is a complete $h$-recursion representation of the torus $N$-point block in the necklace channel.

\subsection{Sphere $N$-point block in the linear channel}

The sphere $N$-point Virasoro conformal block in the linear channel can be obtained as a limit of the torus $(N-1)$-point necklace block, by sending $q_{N-2},q_{N-1}\to 0$. The weights $h_{N-2}$ and $h_{N-1}$ will now be viewed as weights of a pair of external primary operators. This makes it clear that our $h$-recursion relation will involve simultaneous shift of internal weights together with a pair of external weights, which is rather different from the procedure of \cite{1987TMP....73.1088Z}. 

It is nonetheless useful to write the recurrence relation in the sphere linear channel in a set of notations adapted to the Riemann sphere as below. The linear channel conformal block amounts to inserting complete bases of states between successive pairs of external operators, except for the two pairs at the ends, as shown in Figure~\ref{fig:necklacelinear}. This conformal block has been studied in \cite{Alba:2010qc} from the perspective of the AGT relation. Mapping the torus to the annulus by exponentiation, the expansion parameters $q_i$ used in the previous section are related to the positions $z_i$ of the external operators on the complex plane by
\ie
z_1 = 0, z_{N-1} = 1, z_N = \infty,~~z_{i+1}=q_iq_{i+1}...q_{N-3}~~~ \text{for}~ 1\leq i \leq N-3.
\fe
The sphere $N$-point function admits the Virasoro conformal block decomposition
\ie
&\langle O_N(\infty)O_{N-1}(1)O_{N-2}(z_{N-2})...O_2(z_2)O_1(0)\rangle_{S^2}
\\
=&\sum_{(h_1,{\bar h}_1),...,(h_{N-3},{\bar h}_{N-3})}C_{h_1,d_{2},d_1}^{\bar{h}_1,\bar{d}_{2},\bar{d}_{1}}C_{d_N,d_{N-1},h_{N-3}}^{\bar{d}_N,\bar{d}_{N-1},\bar{h}_{N-3}}\left(\prod_{i=1}^{N-4}C_{h_{i+1},d_{i+2},h_{i}}^{\bar{h}_{i+1},\bar{d}_{i+2},\bar{h}_{i}}\right)
\\
&\times z_2^{h_1-d_2-d_1}\bar{z}_2^{\bar{h}_1-\bar{d}_2-\bar{d}_1}\left(\prod_{i=3}^{N-2}z_{i}^{h_{i-1}-d_{i}-h_{i-2}}\bar{z}_i^{\bar{h}_{i-1}-\bar{d}_i-\bar{h}_{i-2}}\right)
\\
&\times F\left(q_1,h_1,,...,q_{N-3},h_{N-3},d_1,...,d_N,c\right)F\left({\bar q}_1,{\bar h}_1,,...,{\bar q}_{N-3},{\bar h}_{N-3},{\bar d}_1,...,{\bar d}_N,c\right),
\fe
where $F(q_i,h_i,d_j,c)$ is the linear channel block
\ie\label{linblk}
&F(q_i,h_i,d_j,c)
\\
=&\sum_{n_1,...,n_{N-3}=0}^{\infty}\left(\prod_{i=1}^{N-3}q_i^{n_i}\right)\sum_{|A_1|=|B_1|=n_1}...\sum_{|A_{N-3}|=|B_{N-3}|=n_{N-3}}\left[\prod_{i=1}^{N-4}\left(G^{n_i}_{h_i}\right)^{A_iB_i}\R(L_{-B_{i+1}}h_{i+1},d_{i+2},L_{-A_{i}}h_{i})\right]
\\
&\times \left(G^{n_{N-3}}_{h_{N-3}}\right)^{A_{N-3}B_{N-3}}\R(d_N,d_{N-1},L_{-A_{N-3}}h_{N-3})\R(L_{-B_1}h_1,d_{2},d_{1}).
\fe
For any $i$ between 1 and $N-3$, we could analytically continue the conformal block in $h_i$, and write a simple pole expansion analogously to (\ref{ipole}), (\ref{necklaceres}),
\ie
F(q_i,h_i,d_j,c)=&U_i+\sum_{1\leq r_is_i}{V_{r_is_i}\over h_i-d_{r_is_i}},
\fe
where the residues are given by
\ie
&V_{r_1s_1}=q_1^{r_1s_1}R_{r_1s_1}(d_{1},h_{2},d_{2},d_{3},c)F(h_1\rightarrow d_{r_1s_1}+r_1s_1),
\\
&V_{r_is_i}=q_i^{r_is_i}R_{r_is_i}(h_{i-1},h_{i+1},d_{i+1},d_{i+2},c)F(h_i\rightarrow d_{r_is_i}+r_is_i),~~~2\leq i\leq N-4,
\\
&V_{r_{N-3}s_{N-3}}=q_{N-3}^{r_{N-3}s_{N-3}}R_{r_{N-3}s_{N-3}}(h_{N-4},d_{N},d_{N-2},d_{N-1},c)F(h_{N-3}\rightarrow d_{r_{N-3}s_{N-3}}+r_{N-3}s_{N-3}).
\fe

To determine the regular part and thereby the full recurrence relation via the large weight limit, it is important to specify how this limit is taken. As in the torus case, we will consider the {\it simultaneous} large $d_1, h_1,\cdots, h_{N-3}, d_N$ limit. In other words, we will define $a_i=h_i-h_1$ for $i=2,...,N-3$ and $e_1=d_1-h_1$, $e_N=d_N-h_1$, and consider the limit $h_1\rightarrow\infty$ with $a_i,e_1,e_N$ held fixed.
Following the same arguments as in Section \ref{hregular}, in (\ref{linblk}) only the terms with the equal internal levels $n_1=...=n_{N-3}$ may survive. However, due to the extra inverse Gram matrix element $\left(G^{n_{N-3}}_{h_{N-3}}\right)^{A_{N-3}B_{N-3}}$, only the internal level zero contribution survives (this can also be understood as effectively sending $q_{N-2}, q_{N-3}$ to zero). Therefore, in this limit, we have simply $F(q_i,h_i,d_j,c)\rightarrow1$.

Combining these results, we obtain the following recursive representation of the sphere $N$-point Virasoro block in the linear channel
\ie\label{linearfinal}
&F(q_i,h_1,a_2,...,a_{N-3},e_1,e_N,c)=1
\\
&+\sum_{r_1s_1\geq1}{q_1^{r_1s_1}R_{r_1s_1}(d_{r_1s_1}+e_1,d_{r_1s_1}+a_2,d_2,d_3,c)\over h_1-d_{r_1s_1}}F(h_1\rightarrow d_{r_1s_1}+r_1s_1, a_i\rightarrow a_i-r_1s_1,e_j\rightarrow e_j-r_1s_1)
\\
&+\sum_{i=2}^{N-4}\sum_{r_is_i\geq1}{q_i^{r_is_i}R_{r_is_i}(d_{r_is_i}-a_i+a_{i-1},d_{r_is_i}-a_i+a_{i+1},d_{i+1},d_{i+2},c)\over h_1+a_i-d_{r_is_i}}F(h_1\rightarrow d_{r_is_i}-a_i,a_i\rightarrow a_i+r_is_i)
\\
&+\sum_{rs\geq1}{q_{N-3}^{rs}R_{rs}(d_{rs}-a_{N-3}+e_N,d_{rs}-a_{N-3}+a_{N-4},d_{N-1},d_{N-2},c)\over h_1+a_{N-3}-d_{rs}}\\
&\times F(h_1\rightarrow d_{rs}-a_{N-3},a_{N-3}\rightarrow a_{N-3}+rs).
\fe
Again, it is important to keep in mind that the shifted blocks on the RHS are functions of $a_i=h_i-h_1$, and thus when viewed as functions of the $h_i$'s, they still contain $h_1$ dependence.

Let us comment that there is another expression for the sphere $N$-point linear channel block in terms of a $q_i$ expansion (which also easily extends to the torus necklace channel) obtained from the AGT relation \cite{Alday:2009aq, Alba:2010qc}. In the language of the latter, such channels include only fundamental, anti-fundamental, or bi-fundamental hypermultiplets, whose Nekrasov instanton partition functions have simplified expressions. 
The instanton partition function gives a combinatorial formula for the Virasoro conformal blocks in these channels. Of course, these expressions should agree with (\ref{necklacefinal}) and (\ref{linearfinal}). This can be verified by showing that the residues and large weight asymptotics agree. It is not hard to check that the simultaneous large weight limit of the combinatorical formula of \cite{Alba:2010qc} is finite. The residues were checked in \cite{Fateev:2009aw,Hadasz:2010xp} for a small number of external operators.

\section{$c$-recursion for all sphere and torus Virasoro conformal blocks}\label{ctorus}

In this section, we derive recursive representation in the central charge $c$ for sphere and torus $N$-point Virasoro conformal blocks in arbitrary channels. The pole structure of the blocks in $c$ is similar to the analytic property in $h$ considered in the previous section: the poles are associated with degenerate Virasoro representations, while the residues are given by appropriate fusion polynomials multiplying the blocks with shifted weights, as will follow from a generic factorization property of 3-point functions of Virasoro descendants. 

The key feature that will allow for the determination of $c$-recursion relations in all channels (in contrast to just the linear and necklace channels in our $h$-recursion relation) will be a drastic simplification in the large $c$ limit. In this limit, the block reduces to the product of the Virasoro vacuum block (i.e. all primaries, both internal and external, are replaced by the identity operator) and a global $SL(2)$ block that captures the contributions of $L_{-1}^n$ descendants of the primaries only. In the sphere case, the vacuum block is just 1, while for the torus, the vacuum block is the Virasoro vacuum character. The global block will be relatively simple to compute.

Throughout this paper we construct Virasoro conformal blocks in terms of $\rho(\xi_3,\xi_2,\xi_1)$, the 3-point function of descendants on the plane. As remarked in section \ref{defvirasoro}, this is natural in the conformal frame where the Riemann surface in question is formed by plumbing together 2-holed discs with $SL(2)$ maps. In describing torus and higher-genus conformal blocks, we could alternatively have made use of $\tilde{\rho}(\xi_3,\xi_2,\xi_1)$, the matrix element of the descendant $\xi_2$ between $\langle \xi_3|$ and $|\xi_1\rangle$ on the cylinder, which would be natural in an alternative conformal frame in which the Riemann surface is formed by plumbing together 1-holed cylinders. While $\tilde\rho(\xi_3, \xi_2, \xi_1)$ can in principle be put in the form $\rho(\xi_3, \xi_2', \xi_1)$ via the exponential map from the cylinder to the plane, the conformally transformed descendant $\xi_2'$ generally differs from $\xi_2$. Different conformal frames not only lead to different parameterizations of the moduli, but also conformal blocks that differ by a conformal anomaly factor (a simple example is the Casimir energy on the cylinder). The simplification at large $c$ mentioned above only holds in the conformal frame defined by the plumbing construction based on 2-holed discs; for this purpose, $\rho(\xi_3,\xi_2,\xi_1)$ rather than $\tilde{\rho}(\xi_3,\xi_2,\xi_1)$ is the appropriate 3-point function building block.

\subsection{Factorization of 3-point functions with degenerate representations and the poles of conformal blocks}\label{genericresidue}

Previously, in our derivation of the $h$-recursive representation of the necklace and linear channel blocks, a key ingredient that allowed for the determination of the polar part of the block was the factorization property of 3-point functions that involve descendants of degenerate primaries (\ref{factorization}) and (\ref{factorization2}). Here we will need a slightly more general set of identities,
\ie\label{Factorization}
\R(L_{-N}\X_{rs},L_{-M}\n_2,L_{-P}\n_3|1)&=\R(L_{-N}\n_{d_{rs}+rs},L_{-M}\n_2,L_{-P}\n_3|1)\R(\X_{rs},\n_2,\n_3|1)
\\
\R(L_{-N}\n_1,L_{-M}\X_{rs},L_{-P}\n_3|1)&=\R(L_{-N}\n_1,L_{-M}\n_{d_{rs}+rs},L_{-P}\n_3|1)\R(\n_1,\X_{rs},\n_3|1)
\\
\R(L_{-N}\n_1,L_{-M}\n_2,L_{-P}\X_{rs}|1)&=\R(L_{-N}\n_1,L_{-M}\n_2,L_{-P}\n_{d_{rs}+rs}|1)\R(\n_1,\n_2,\X_{rs}|1)
\\
\R(L_{-N}\X_{rs},L_{-M}\n_2,L_{-P}\X_{rs}|1)&=\R(L_{-N}\n_{d_{rs}+rs},L_{-M}\n_2,L_{-P}\n_{d_{rs}+rs}|1)\R(\X_{rs},\n_2,\X_{rs}|1).
\fe
We remind the reader that $\chi_{rs}$ is the level $rs$ null descendant of a primary of weight $d_{rs}$, of the form (\ref{null}). (\ref{Factorization}) follows from Ward identities and the property that $\chi_{rs}$ behaves like a Virasoro primary, as explained in more detail in Appendix~\ref{app:3pt}. 

On the RHS of (\ref{Factorization}), the first factors will lead to the recursive representation, as they contribute to new conformal blocks with shifted internal weight $d_{rs}+rs$. The second factors are fusion polynomials $P_{rs}^c$ (\ref{app:fusionpolynomials}). Together, (\ref{Factorization}) will determine the residue of a Virasoro conformal block on its poles either at a degenerate value of an internal weight, $h_i\rightarrow d_{rs}$, or at a value of the central charge $c\rightarrow c_{rs}(h_i)$ such that an internal weight $h_i$ becomes that of a degenerate Virasoro representation. This statement applies to any $N$-point, genus $g$ Virasoro block in any given channel, as will become clear in the next section. 

Consider for example the sphere 6-point block shown in Figure~\ref{fig:BlockExamples}, which we refer to as the ``trifundamental" channel block.\footnote{The terminology comes from the corresponding quiver theory in the context of the AGT relation, which involves a trifundamental hypermultiplet \cite{Hollands:2011zc}.} We may build the 6-punctured sphere by connecting three 2-punctured discs and a single two-holed disc through the following plumbing construction. Consider the 2-punctured and 2-holed discs
\ie
& D_i = \{z_i\in\mathbb{C}: ~|z_i|<r_i, ~z_i\not=0,1\}, ~~~i=1,2,3
\\
& D_4 = \{z_4 \in \mathbb{C}:~\tilde r_1< |z_4|< \tilde r_3,~ |z_4-1|>\tilde r_2\}.
\fe

\begin{figure}[h!]
\centering
\subfloat{\includegraphics[width=.59\textwidth]{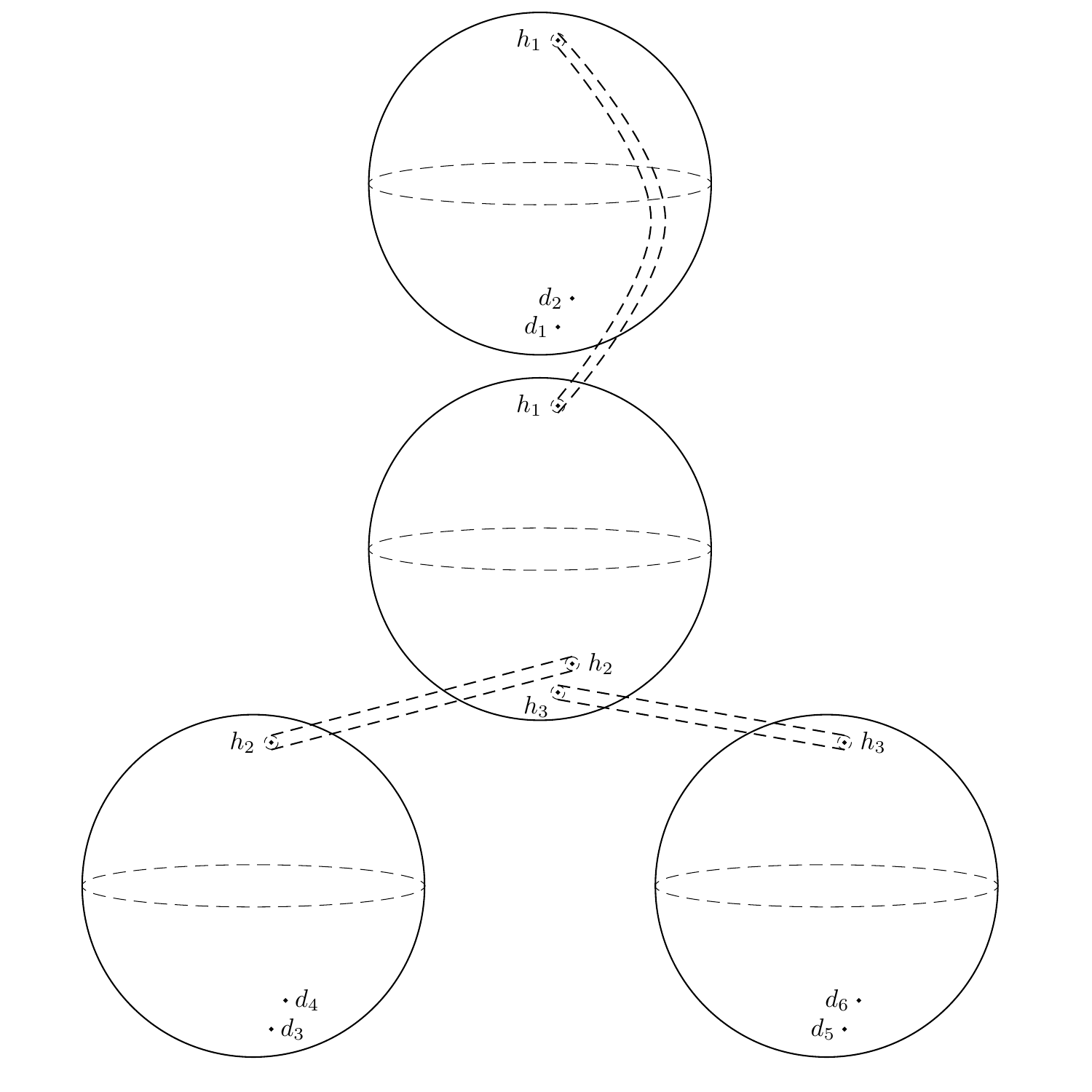}}
\caption{Plumbing construction for sphere 6-point conformal block in the trifundamental channel.}\label{fig:threepunctured}
\end{figure}

We glue each boundary component of $D_4$ with the boundary of $D_i$, $i=1,2,3$, via the $SL(2)$ maps
\ie
& |z_4|=\tilde r_1:~~ z_4 = q_1 z_1,~~ |q_1| = {\tilde r_1\over r_1},
\\
& |z_4-1|=\tilde r_2:~~ z_4-1 = q_2 z_2,~~ |q_2| = {\tilde r_2\over r_2},
\\
& |z_4|=\tilde r_3:~~ z_4 = {1\over q_3 z_3},~~ |q_3| = {1\over \tilde r_3 r_3}.
\fe
The result of the plumbing construction is a Riemann sphere with 6 punctures at
\ie
0,~~ q_1,~~ 1,~~ 1+q_2,~~ \infty, ~~ {1\over q_3}.
\fe
The 6 external vertex operators will be inserted at these 6 points, parameterized by the plumbing parameters $q_1, q_2, q_3$. Note that $q_1, q_2, q_3$ are {\it not} on equal footing.
In such a parameterization, the Virasoro block is given by
\ie\label{trifund}
F(q_i,h_i,d_j,c)=&\sum_{N,M,P,Q,A,B}q_1^{|N|}q_2^{|P|}q_3^{|A|}\R(L_{-N}h_1,d_2,d_1)\R(L_{-P}h_2,d_4,d_3)\R(L_{-A}h_3,d_6,d_5)
\\
&\times\R(L_{-M}h_1,L_{-Q}h_2,L_{-B}h_3)\left(G^{|N|}_{h_1}\right)^{NM}\left(G^{|P|}_{h_2}\right)^{PQ}\left(G^{|A|}_{h_3}\right)^{AB}.
\fe
Here the summation is over integer partitions $N,M,P,Q,A,B$, with $|N|=|M|$, $|P|=|Q|$, $|A|=|B|$, that label Virasoro descendants.

A simple pole expansion of this conformal block in one of the weights, say $h_1$, takes the form
\ie
F(q_i,h_i,d_j,c)=U_1+\sum_{rs\geq1}{q_1^{rs}A_{rs}^cP^{rs}_c\begin{bmatrix}d_1\\d_2\end{bmatrix}P^{rs}_c\begin{bmatrix}h_3\\h_2\end{bmatrix}\over h_1-d_{rs}}F(h_1\rightarrow d_{rs}+rs),
\fe
where the residue is readily determined using the factorization property (\ref{Factorization}) as before. The $h_1$-regular part $U_1$ is more complicated. Instead of trying to determine $U_1$ directly, we can inspect similar polar terms in $h_2$ and $h_3$, and write a simple pole expansion in the central charge $c$ using (\ref{hcresidue}), of the form
\ie\label{trifundc}
& F(q_i,h_i,d_j,c)
\\
&=U_c+\sum_{r\geq2,s\geq1} \left[ -{\partial c_{rs}(h_1) \over\partial h_1}  \right] {q_1^{rs}A^{c_{rs}(h_1)}_{rs}P^{rs}_{c_{rs}(h_1)}\begin{bmatrix}d_1\\d_2\end{bmatrix}P^{rs}_{c_{rs}(h_1)}\begin{bmatrix}h_3\\h_2\end{bmatrix}\over c-c_{rs}(h_1)}F(h_1\rightarrow h_1+rs,c\rightarrow c_{rs}(h_1))
\\
&~~~~+\sum_{r\geq2,s\geq1} \left[-{\partial c_{rs}(h_2)\over\partial h_2} \right] {q_2^{rs}A^{c_{rs}(h_2)}_{rs}P^{rs}_{c_{rs}(h_2)}\begin{bmatrix}d_3\\d_4\end{bmatrix}P^{rs}_{c_{rs}(h_2)}\begin{bmatrix}h_3\\h_1\end{bmatrix}\over c-c_{rs}(h_2)}F(h_2\rightarrow h_2+rs,c\rightarrow c_{rs}(h_2))
\\
&~~~~+\sum_{r\geq2,s\geq1} \left[ -{\partial c_{rs}(h_3)\over\partial h_3} \right] {q_3^{rs}A^{c_{rs}(h_3)}_{rs}P^{rs}_{c_{rs}(h_3)}\begin{bmatrix}d_5\\d_6\end{bmatrix}P^{rs}_{c_{rs}(h_3)}\begin{bmatrix}h_1\\h_2\end{bmatrix}\over c-c_{rs}(h_3)}F(h_3\rightarrow h_3+rs,c\rightarrow c_{rs}(h_3)).
\fe
Now the $c$-regular part $U_c$ is the only term that survives in the large-$c$ limit. This will be analyzed next.

\subsection{Large $c$, fixed $h_i$ limit of Virasoro conformal blocks}\label{finitec}

In the previous subsection, we have seen that the factorization property (\ref{Factorization}) fixes the polar part of the recursive representation of an arbitrary Virasoro conformal block, and the problem reduces to determining the large $c$ limit of the block, such as $U_c$ in the case of the trifundamental block (\ref{trifundc}). We now show that a general Virasoro conformal block built out of descendant 3-point functions of the form $\R(L_{-A}h_i,L_{-B}h_j,L_{-C}h_k)$ ($A,B,C$ stand for integer partitions) and inverse Gram matrices remains finite in the $c\to\infty$ limit (rather than growing with $c$). Furthermore, it will turn out that the large $c$ limit of a Virasoro conformal block factorizes into the product of the large $c$ limit of the vacuum block (defined by setting all internal and external representations to the vacuum) and the global $SL(2)$ conformal block. 

Note that the construction of the Virasoro block using descendant 3-point functions and inverse Gram matrices amounts to a plumbing construction based on gluing together 2-holed discs via the inversion map, which specifies the coordinate charts for the punctured Riemann surface as well as the conformal frame of the conformal block. We have already seen such an example in (\ref{trifund}).
In the case of higher genus conformal blocks, this choice of conformal frame fixes the conformal anomaly in such a way that the blocks are $c$-independent to leading order.

We illustrate the large $c$ factorization property by considering the genus two Virasoro conformal block in the channel where the genus two Riemann surface is formed by plumbing together a pair of 2-holed discs. This conformal block takes the form\footnote{This object was referred to as the ``naive" conformal block in \cite{Hollands:2011zc}. 
}
\ie\label{genus2}
F(h_1,h_2,h_3,c)=&\sum_{|A|=|B|, \,|C|=|D|, \,|E|=|F|}q_1^{|A|}q_2^{|C|}q_3^{|E|}G_{h_1}^{AB}G_{h_2}^{CD}G_{h_3}^{EF}\\
&\times\R(L_{-A}h_1,L_{-C}h_2,L_{-E}h_3)\R(L_{-B}h_1,L_{-D}h_2,L_{-F}h_3).
\fe
The strategy here closely parallels that of Section \ref{hregular}, with slight modifications. We begin with a basis of level $n$ descendants of a primary $|h\rangle$ of the form $L_{-A}|h\rangle$, where $A$ is a partition of the integer $n = |A|$ in descending order. We will denote by $\langle A\rangle$ the number of non-$L_{-1}$ Virasoro generators in $L_{-A}$. Note that in the large $c$ limit with $h$ fixed, the inner product $\langle h| L_{-A}^\dagger L_{-B}|h\rangle$ scales like $c^{\langle A\rangle}$ for $A=B$, no faster than $c^{\langle A\rangle-1}$ for $\langle A\rangle=\langle B\rangle$, $A\not=B$, and no faster than $c^{{\rm min}(\langle A\rangle,\langle B\rangle)}$ for $\langle A\rangle\not=\langle B\rangle$. We can thus construct via the Gram-Schmidt process an orthogonal basis of the form
\ie\label{corthob}
\cL_{-A} |h\rangle = L_{-A}|h\rangle + \sum_{|B|=n,~ \langle B\rangle\leq \langle A\rangle,~B\not=A} g^A_B(c,h)L_{-B} |h\rangle,
\fe
such that 
\ie\label{gscaling}
& g^A_B(c,h) \sim {\cal O}(c^{-1}),~~~ \langle B\rangle=\langle A\rangle,~B\not=A;
\\
& g^A_B(c,h) \sim {\cal O}(c^0),~~~ \langle B\rangle<\langle A\rangle,
\fe
in the large $c$, fixed $h$ limit. The norm of the basis state $\cL_{-A}|h\rangle$ scales like
\ie
\langle h| \cL_{-A}^\dagger \cL_{-A} |h\rangle \sim c^{\langle A\rangle}.
\fe
(\ref{genus2}) can now be written as
\ie\label{abcf}
F(h_1,h_2,h_3,c)=\sum_{A,B,C}q_1^{|A|}q_2^{|B|}q_3^{|C|}{\R(\cL_{-A} h_1,\cL_{-B} h_2, \cL_{-C} h_3)^2\over\langle h_1| \cL_{-A}^\dagger \cL_{-A} |h_1\rangle\langle h_2| \cL_{-B}^\dagger \cL_{-B} |h_2\rangle\langle h_3| \cL_{-C}^\dagger \cL_{-C} |h_3\rangle}.
\fe
By construction of (\ref{corthob}), the three-point function $\R(\cL_{-A} h_1,\cL_{-B} h_2, \cL_{-C} h_3)$ scales with $c$ no faster than $c^{\langle A\rangle+\langle B\rangle+\langle C\rangle\over2}$. Therefore, (\ref{genus2}) is finite in the $c\to \infty$ limit. Moreover, in this limit the only surviving contribution to $\R(\cL_{-A} h_1,\cL_{-B} h_2, \cL_{-C} h_3)$ comes from the $L_{-A}$ term in $\cL_{-A}$ (\ref{corthob}), i.e.
\ie\label{largec3pt}
\R(\cL_{-A} h_1,\cL_{-B} h_2, \cL_{-C} h_3)\rightarrow\R(L_{-A} h_1,L_{-B} h_2, L_{-C} h_3).
\fe
To prove the large $c$ factorization into the vacuum Virasoro block and the global $SL(2)$ block, we write Virasoro chains as $L_{-A}=L_{-A'}L_{-1}^{k_A}$, where $L_{-A'}$ does not include any $L_{-1}$ generators (by convention, $A'$ and $A$ are both integer partitions in descending order). The RHS of (\ref{largec3pt}) is now written as $\R(L_{-A'}L_{-1}^{k_A} h_1,L_{-B'}L_{-1}^{k_B} h_2, L_{-C'}L_{-1}^{k_C} h_3)$. To leading order in the large $c$ limit, the (non-$L_{-1}$) Virasoro generators in $L_{-A'}, L_{-B'}, L_{-C'}$ must be contracted pairwise via the Ward identities. In particular, the dependence on the weights $h_i$ is suppressed by $h_i/c$ relative to the leading order scaling $c^{\langle A\rangle+\langle B\rangle+\langle C\rangle\over 2}$ (when $\langle A\rangle+\langle B\rangle+\langle C\rangle$ is even and the pairwise contraction is available).
What remains is the 3-point function of $L_{-1}$ descendants. Thus, in the large $c$ limit we can replace
\ie\label{largec3ptfactorize}
\R(\cL_{-A} h_1,\cL_{-B} h_2, \cL_{-C} h_3)\rightarrow\R(L_{-A'}\n_{0},L_{-B'}\n_{0}, L_{-C'}\n_{0})\R(L_{-1}^{k_A} h_1,L_{-1}^{k_B}h_2, L_{-1}^{k_C} h_3),
\fe
where $\n_0$ is the vacuum primary. Note that if the 3-point function of vacuum descendants on the RHS of (\ref{largec3ptfactorize}) vanishes, the contribution to the conformal block also vanishes in the $c\to \infty$ limit, due to the factorization property of the 2-point function 
\ie
\langle h| \cL_{-A}^\dagger \cL_{-A} |h\rangle\rightarrow \langle \n_0|L_{-A'}^\dagger L_{-A'}|\n_0\rangle\langle h|L_{-1}^{k_A\dagger}L_{-1}^{k_A}|h\rangle.
\fe
Rewriting the summation over partitions $A,B,C$ in (\ref{abcf}) in terms of $(A', k_A)$, $(B', k_B)$, $(C', k_C)$, where $A', B', C'$ involve only non-$L_{-1}$ generators, and $k_A, k_B, k_C$ counts the length of the $L_{-1}$ chains, we arrive at the large $c$ limit
\ie\label{genus2largec}
\lim_{c\rightarrow\infty}F(h_1,h_2,h_3)=& \lim_{c\rightarrow\infty}\sum_{A',B',C'}q_1^{|A'|}q_2^{|B'|}q_3^{|C'|}{\R(L_{-A'} \n_0,L_{-B'} \n_0, L_{-C'} \n_0)^2\over\langle \n_0| L_{-A'}^\dagger L_{-A'} |\n_0\rangle\langle \n_0| L_{-B'}^\dagger L_{-B'} |\n_0\rangle\langle \n_0| L_{-C'}^\dagger L_{-C'} |\n_0\rangle} 
\\
&\times \sum_{k_1,k_2,k_3\geq0}q_1^{k_1}q_2^{k_2}q_3^{k_3}{\R(L_{-1}^{k_1} h_1,L_{-1}^{k_2}h_2, L_{-1}^{k_3} h_3)^2\over \langle h_1|L_{-1}^{k_1\dagger}L_{-1}^{k_1}|h_1\rangle\langle h_2|L_{-1}^{k_2\dagger}L_{-1}^{k_2}|h_2\rangle\langle h_3|L_{-1}^{k_3\dagger}L_{-1}^{k_3}|h_3\rangle}.
\fe
The first factor on the RHS is the large $c$ limit of the vacuum block (note that the vacuum is annihilated by $L_{-1}$), while the second factor is the global $SL(2)$ conformal block which by definition is independent of the central charge.

Clearly, the above proof can be straightforwardly extended to any Virasoro conformal blocks built from contracting 3-point functions of descendants with inverse Gram matrices, as the argument was simply based on power counting in the large $c$ limit. 
Note that the vacuum Virasoro block on the sphere is equal to 1, and vacuum Virasoro block on the torus (in any channel) is equal to the vacuum Virasoro character. Thus, the large $c$ limit for any $N$-point sphere or torus Virasoro conformal block in any channel $\cC$ (corresponding to a pair-of-pants decomposition of the $N$-punctured Riemann surface) is given by
\ie\label{climitsphere}
&\lim_{c\rightarrow\infty}(\text{sphere Virasoro block in channel $\cC$})=(\text{sphere global $SL(2)$ block in channel $\cC$})
\\
&\lim_{c\rightarrow\infty}(\text{torus Virasoro block in channel $\cC$})\\
&=(\text{vacuum Virasoro character})\times(\text{torus global $SL(2)$ block in channel $\cC$}).
\fe
Together with the residue structure of the $c$-polar part discussed in the previous section, we obtain a complete $c$-recursive representation for any $N$-point sphere and torus Virasoro conformal block. In the next two subsections, we will give the explicit formulae in several examples.

An analogous large $c$ factorization property holds for higher genus Virasoro conformal blocks as well, provided that we define the latter in the appropriate conformal frame, based on plumbing together 2-holed discs. This will be discussed in section \ref{highergenussection}.

\subsection{Global $SL(2)$ blocks}

Here we briefly describe the evaluation of global $SL(2)$ blocks. Consider as an example the sphere 6-point block in the trifundamental channel (\ref{trifund}), defined in terms of the plumbing parameters $q_1, q_2, q_3$. Its corresponding global block reads
\ie\label{trifundglobal}
G(q_i,h_i,d_j,c)&=\sum_{i,j,k=0}^\infty q_1^iq_2^jq_3^k\R(L_{-1}^ih_1,d_2,d_1)\R(L_{-1}^jh_2,d_4,d_3)\R(L_{-1}^kh_3,d_6,d_5)
\\
&~~~\times{\R(L_{-1}^ih_1,L_{-1}^jh_2,L_{-1}^kh_3)\over \langle h_1|L_1^iL_{-1}^i|h_1\rangle\langle h_2|L_1^jL_{-1}^j|h_2\rangle\langle h_3|L_1^kL_{-1}^k|h_3\rangle}.
\fe
The global block is generally simple enough to evaluate in closed form. For instance,
\ie
\R(L^i_{-1}h_1,d_2,d_1)=(h_1+d_2-d_1)_i,
\fe
where $(a)_i$ is the Pochhammer symbol. The most general 3-point function of $L_{-1}$ descendants is
\ie\label{genericglobal}
\R(L_{-1}^ih_1,L_{-1}^jh_2,L_{-1}^kh_3)=(h_1+i-h_2-j+1-h_3-k)_js_{ik}(h_1,h_2,h_3),
\fe
where we have defined $s_{ik}(h_1,h_2,h_3)=\R(L_{-1}^ih_1,h_2,L_{-1}^kh_3)$. It has a known closed form expression \cite{Alkalaev:2015fbw}
\ie
s_{km}(h_1,h_2,h_3)&=\sum_{p=0}^{\text{min}(m,k)}{k!\over p!(k-p)!}(2h_3+m-1)^{(p)}m^{(p)}
\\
&~~~\times (h_3+h_2-h_1)_{m-p}(h_1+h_2-h_3+p-m)_{k-p},
\fe
where $(a)^{(p)}=a(a-1)...(a-p+1)$ is the descending Pochhammer symbol. The inverse norms in (\ref{trifundglobal}) are given by
\ie
{1\over\langle h|L_1^nL_{-1}^n|h\rangle}={1\over n!(2h)_n}.
\fe
Combining these, we arrive at the following closed form expression for the trifundamental global block 
\ie
G=&\sum_{i,j,k=0}^\infty q_1^iq_2^jq_3^k(h_1+d_2-d_1)_i(h_2+d_4-d_3)_j(h_3+d_6-d_5)_k
\\
&\times{(h_1+i-h_2-j+1-h_3-k)_js_{ik}(h_1,h_2,h_3)\over i!j!k!(2h_1)_i(2h_2)_j(2h_3)_k}.
\fe
The extension of such results to any global block is evident. Let us note that for a given channel of an $N$-point, genus $g$ conformal block, based on a pair-of-pants decomposition, the global $SL(2)$ block is only {\it defined} in the plumbing construction based on 2-holed discs glued together via $SL(2)$ maps.

\subsection{Examples of $c$-recursive representations}\label{sec:cRecursionExamples}

\subsubsection{Sphere 6-point block in the trifundamental channel}

Our first nontrivial example is the sphere 6-point block in the trifundamental channel (\ref{trifund}). (Note that the $h$-recursive representation given in the previous section is not available in this channel.) Combining the large $c$ limit and the polar structure determined earlier, we have the following $c$-recursion formula
\ie\label{sphtri}
& F(q_i,h_i,d_j,c)
\\
&=U_c+\sum_{r\geq2,s\geq1} \left[ -{\partial c_{rs}(h_1)\over\partial h_1}\right] {q_1^{rs}A^{c_{rs}(h_1)}_{rs}P^{rs}_{c_{rs}(h_1)}\begin{bmatrix}d_1\\d_2\end{bmatrix}P^{rs}_{c_{rs}(h_1)}\begin{bmatrix}h_3\\h_2\end{bmatrix}\over c-c_{rs}(h_1)}F(h_1\rightarrow h_1+rs,c\rightarrow c_{rs}(h_1))
\\
&~~~+\sum_{r\geq2,s\geq1} \left[ -{\partial c_{rs}(h_2)\over\partial h_2}\right] {q_2^{rs}A^{c_{rs}(h_2)}_{rs}P^{rs}_{c_{rs}(h_2)}\begin{bmatrix}d_3\\d_4\end{bmatrix}P^{rs}_{c_{rs}(h_2)}\begin{bmatrix}h_3\\h_1\end{bmatrix}\over c-c_{rs}(h_2)}F(h_2\rightarrow h_2+rs,c\rightarrow c_{rs}(h_2))
\\
&~~~+\sum_{r\geq2,s\geq1} \left[ -{\partial c_{rs}(h_3)\over\partial h_3}\right] {q_3^{rs}A^{c_{rs}(h_3)}_{rs}P^{rs}_{c_{rs}(h_3)}\begin{bmatrix}d_5\\d_6\end{bmatrix}P^{rs}_{c_{rs}(h_3)}\begin{bmatrix}h_1\\h_2\end{bmatrix}\over c-c_{rs}(h_3)}F(h_3\rightarrow h_3+rs,c\rightarrow c_{rs}(h_3)),
\fe
with
\ie
U_c=&\sum_{i,j,k=0}^\infty q_1^iq_2^jq_3^k(h_1+d_2-d_1)_i(h_2+d_4-d_3)_j(h_3+d_6-d_5)_k
\\
&\times{(h_1+i-h_2-j+1-h_3-k)_js_{ik}(h_1,h_2,h_3)\over i!j!k!(2h_1)_i(2h_2)_j(2h_3)_k}.
\fe

\subsubsection{Torus 1-point block}

Our next example is the torus 1-point block, which was already considered in \cite{Hadasz:2009db,Alkalaev:2016fok}. Properties of this block were used to derive an asymptotic formula for the average value of heavy-heavy-light OPE coefficients from modular covariance of the torus 1-point function in \cite{Kraus:2016nwo}. The block is given by
\ie\label{t1}
F(q,h,d,c)=\sum_{|N|=|M|=n}q^n\R(L_{-N}h,d,L_{-M}h)(G^n_h)^{NM},
\fe
where $q=e^{2\pi i \tau}$, $\tau$ being the modulus of the torus. Our conformal frame is defined by identifying the inner and outer boundaries of the annulus via the rescaling $z\mapsto q^{-1} z$ on the complex plane, and thus the Casimir energy factor $q^{-{c\over 24}}$ is absent. This distinction is rather minor in the present example, but will be important in more complicated examples to be discussed later.

The recursive representation of the torus 1-point block in the internal weight $h$ reads \cite{Hadasz:2009db}
\ie\label{t1h}
& F(q,h,d,c) = \left[\prod_{n=1}^\infty {1\over 1-q^n}\right]f(q,h,d,c),
\\
& f(q,h,d,c)=1+\sum_{rs\geq1}{A_{rs}^cP^{rs}_c\begin{bmatrix}d\\d_{rs}+rs\end{bmatrix}P^{rs}_c\begin{bmatrix}d\\d_{rs}\end{bmatrix}\over h-d_{rs}}f(h\rightarrow d_{rs}+rs) .
\fe
Note that here we encounter a 3-point function involving a pair of null states $\chi_{rs}$, resulting in the product of two fusion polynomials that involve the weight $d_{rs}$ and $d_{rs}+rs$ respectively.
The corresponding global $SL(2)$ block is \cite{Hadasz:2009db,Alkalaev:2016fok}
\ie\label{eq:Torus1PtGlobal}
g(q,h,d)={1\over1-q}~{}_2F_1\left(d,1-d;2h;{q\over q-1}\right).
\fe
As originally observed in \cite{Alkalaev:2016fok}, the large $c$ limit of the torus 1-point block reduces to the product of the vacuum Virasoro character with the global block,
\ie\label{t1c}
\lim_{c\to\infty}F(q,h,d,c) =& \left[\prod_{n=2}^\infty{1\over 1-q^n}\right]g(q,h,d)= \left[\prod_{n=1}^\infty{1\over 1-q^n}\right]{}_2F_1\left(d,1-d;2h;{q\over q-1}\right).
\fe
We arrive at the following $c$-recursive representation, in agreement with \cite{Alkalaev:2016fok}
\ie
&F(q,h,d,c)=\left[\prod_{n=1}^\infty{1\over 1-q^n}\right]{}_2F_1\left(d,1-d;2h;{q\over q-1}\right)
\\
&~~+\sum_{r\geq2,s\geq1} \left[ -{\partial c_{rs}(h)\over\partial h}\right] {q^{rs}A^{c_{rs}(h)}_{rs} P^{rs}_{c_{rs}(h)}\begin{bmatrix}d\\h+rs\end{bmatrix}P^{rs}_{c_{rs}(h)}\begin{bmatrix}d\\h \end{bmatrix} \over c-c_{rs}(h)}
\\
&~~~~~~~~~~~~~~~\times F(h\rightarrow h+rs,c\rightarrow c_{rs}(h)),
\fe
where we have used that $d_{rs}(c_{rs}(h)) = h$.

\subsubsection{Torus 2-point block in the OPE channel}

\begin{figure}[h!]
\centering
\subfloat{\includegraphics[width=.8\textwidth]{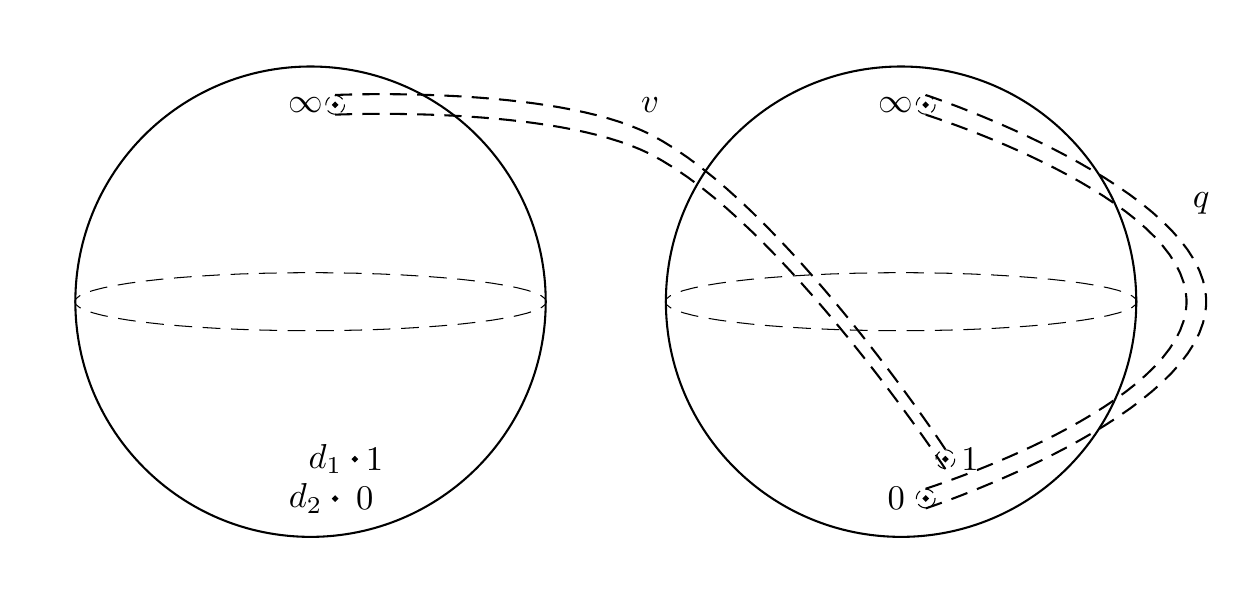}}
\caption{The plumbing construction of the torus 2-point function in the OPE channel.}\label{fig:t2ope}
\end{figure}

The last example is the torus 2-point conformal block in the OPE channel, that is, two external vertex operators fusing into one that is inserted on the torus. Our conformal frame is defined by the plumbing construction illustrated in Figure \ref{fig:t2ope}.

We begin with a 2-punctured disc and a 2-holed disc,
\ie
& D_1 = \{ w\in \mathbb{C}:~ |w|<r_1,~w\not=0,1\},
\\
& D_2 = \{ u \in \mathbb{C}:~ \epsilon<|u|<{\epsilon\over |q|},~ |u-1|>r_2\}.
\fe
The $SL(2)$ gluing maps identify
\ie
& |u-1|=r_2: ~~u-1 = vw,~|v|= {r_2\over r_1},~~{\rm and}
\\
& u \sim q u.
\fe
The result of the plumbing construction is the annulus on the $u$ plane with the identification $u\sim qu$ and two vertex operators inserted at $u=1$ and $u=1+v$. In terms of the parameters $q_1, q_2$ previously used for the necklace channel, we have
\ie
q = q_1 q_2,~~~ v = {1-q_2\over q_2}.
\fe

The Virasoro conformal block in this frame is given by
\ie
F(q,h_1,v,h_2,d_1,d_2,c)=\sum_{N,M,P,Q}q^{|N|}v^{|P|}\left(G_{h_1}^{|N|} \right)^{NM}\R(L_{-N}h_1,L_{-P}h_2,L_{-M}h_1)\left(G_{h_2}^{|P|} \right)^{PQ}\R(L_{-Q}h_2,d_1,d_2).
\fe
It is important that $\R$ is defined as the 3-point function of descendants on the plane (as opposed to on the cylinder), as is clear from the above plumbing construction. The $c$-recursive representation takes the form
\ie
F&=U_c+\sum_{r\geq2,s\geq1} \left[-{\partial c_{rs}(h_1)\over\partial h_1}\right] {q^{rs}A^{c_{rs}(h_1)}_{rs} P^{rs}_{c_{rs}(h_1)}\begin{bmatrix}h_2\\h_1+rs\end{bmatrix}P^{rs}_{c_{rs}(h_1)}\begin{bmatrix}h_2\\h_1\end{bmatrix}\over c-c_{rs}(h_1)}
\\
&~~~~~~~~~~~~~~~~~~~~\times F(h_1\rightarrow h_1+rs,c\rightarrow c_{rs}(h_1))
\\
&+\sum_{r\geq2,s\geq1} \left[ -{\partial c_{rs}(h_2)\over\partial h_2} \right] {v^{rs}A^{c_{rs}(h_2)}_{rs}P^{rs}_{c_{rs}(h_2)}\begin{bmatrix}h_1\\h_1\end{bmatrix}P^{rs}_{c_{rs}(h_2)}\begin{bmatrix}d_2\\d_1\end{bmatrix}\over c-c_{rs}(h_2)}F(h_2\rightarrow h_2+rs,c\rightarrow c_{rs}(h_2)).
\fe
The $c$-regular part $U_c$ is again $c$-independent, and is given by the product of the torus vacuum character and the global block,
\ie
U_c=\left[\prod_{n=2}^\infty{1\over 1-q^n}\right]\sum_{i,j\geq0}q^iv^j{(1-h_2-j)_j s_{ii}(h_1,h_2,h_1)(h_2+d_1-d_2)_j\over i!(2h_1)_i~ j!(2h_2)_j}.
\fe

\section{Generalization to higher genus}\label{highergenussection}

\begin{figure}[h!]
\centering
\subfloat{\includegraphics[width=.9\textwidth]{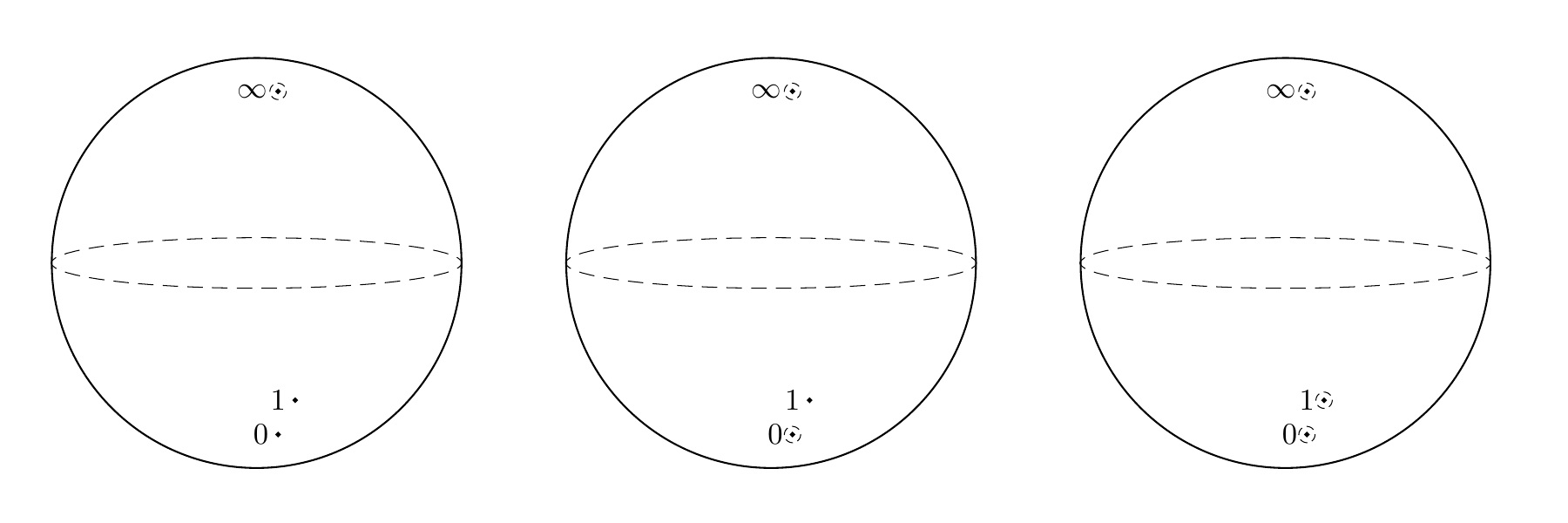}}
\caption{Building blocks for the plumbing construction. Discs with 2 punctures (left), 1 hole and 1 puncture (middle), and 2 holes (right). Here we describe them using spheres with at least one hole around $\infty$, which are $SL(2)$-equivalent to discs with the appropriate number of holes and punctures.}\label{fig:discs}
\end{figure}

We now describe the extension of $c$-recursive representation to $N$-point Virasoro conformal blocks on arbitrary higher genus Riemann surfaces in an arbitrary channel. The $N$-punctured genus $g$ Riemann surface will be constructed by plumbing together $2g-2+N$ discs with either 2 holes, 1 hole and 1 puncture, or 2 punctures as illustrated in Figure \ref{fig:discs}. For instance, a 2-holed disc is the domain
\ie
D=\{z\in \mathbb{C}:~ |z|>r_1, ~|z-1|> r_2, ~|z|< r_3\}.
\fe
Boundary components of the holed/punctured discs will be identified pairwise using $3g-3+N$ $SL(2,\mathbb{C})$ M\"obius maps. For instance, we may glue the inner boundary $|z|=r_1$ of a 2-holed disc $D$ with the outer boundary $|\tilde z|=\tilde r_3$ of another 2-holed disc $\tilde D$ via $\tilde z = z /q$. The moduli of the $N$-punctured genus $g$ Riemann surface will be parameterized by $3g-3+N$ plumbing parameters $q_i$.

\begin{figure}[h!]
\subfloat{\includegraphics[width=.54\textwidth]{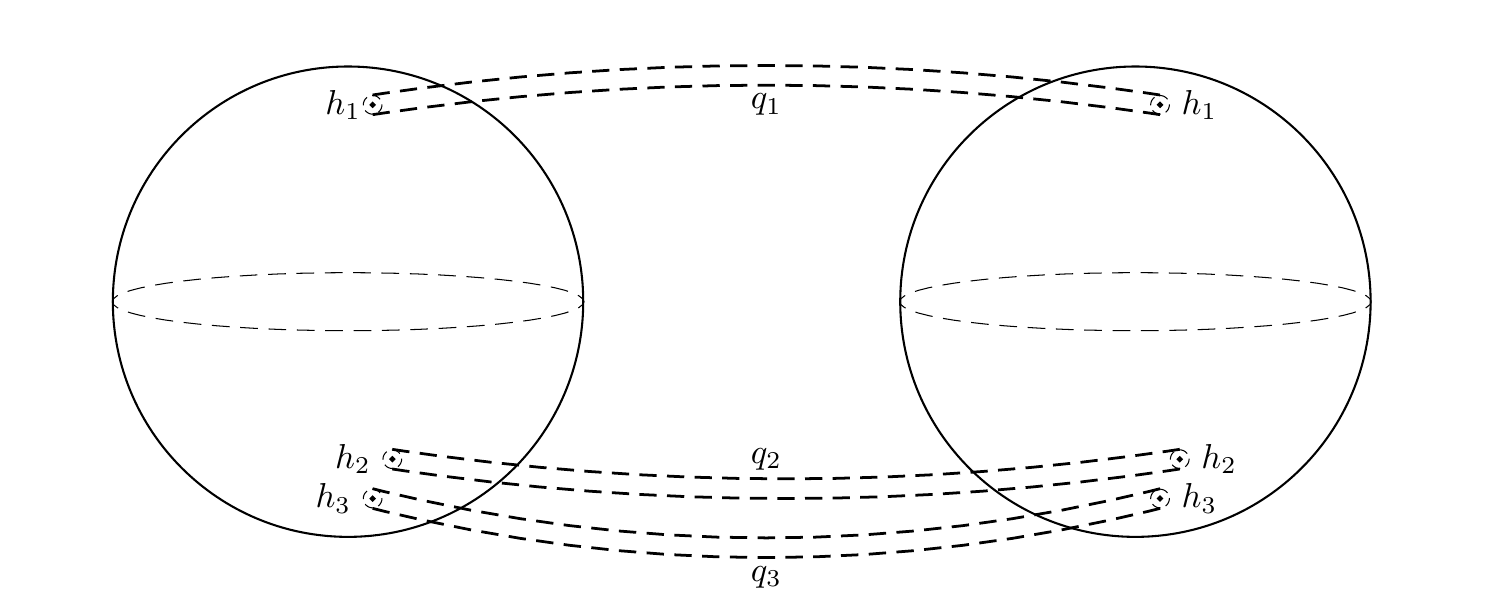}}~
\subfloat{\includegraphics[width=.45\textwidth]{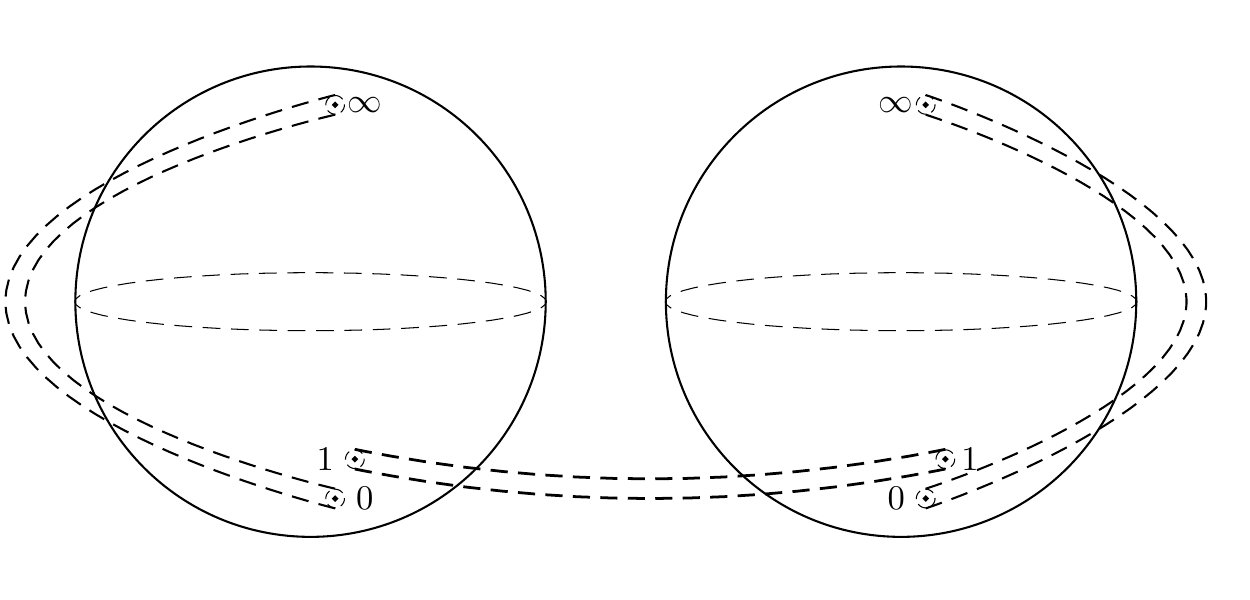}}
\caption{The plumbing construction for a genus two Riemann surface in two channels.}\label{fig:higherg}
\end{figure}

The plumbing construction not only gives a parameterization of the moduli, but also specifies the conformal frame in which the Virasoro conformal block is defined. As already mentioned, this is a particularly convenient frame for the $c$-recursive representation, because (1) the Virasoro conformal block remains finite in the $c\to \infty$ limit in this frame, and (2) the global $SL(2)$ block is naturally defined in this frame since only $SL(2)$ gluing maps are involved.

To build the Virasoro block, we begin with 3-point functions $\R$ of Virasoro descendants inserted at $z=0,1,\infty$ on the plane, associated with each holed/punctured disc. A puncture corresponds to an external primary, while a hole corresponds to an internal descendant of the form $L_{-A}\n_h$. Each gluing map in the plumbing construction amounts to contracting a pair of descendants from two $\rho$'s, say of primary weight $h$ and level $N$, with the inverse Gram matrix, multiplied by a power of the plumbing parameter, $q^{N}$ (by convention, we have separated $q^h$ as an overall prefactor that multiplies the conformal block). We have already seen this through a number of examples:
for instance, the sphere 6-point block in the trifundamental channel (\ref{trifund}) corresponds to Figure \ref{fig:threepunctured}; the genus two conformal block corresponding to the left figure of Figure \ref{fig:higherg} was considered in (\ref{genus2}).

As described in Section \ref{genericresidue}, the factorization property of descendant 3-point functions $\R$ involving null states leads to the determination of the residues of the conformal block at its poles either in one of the internal weights or in the central charge. For instance, the genus two block (\ref{genus2}) has the simple pole expansion in one of its internal weights $h_1$,
\ie\label{hones}
F=U_1+\sum_{rs\geq1}{q_1^{rs}A_{rs}\left(P^{rs}\begin{bmatrix}h_3\\h_2\end{bmatrix}\right)^2\over h_1-d_{rs}}F(h_1\rightarrow d_{rs}+rs),
\fe
where $U_1$ is regular in $h_1$. Similar results of course hold for the simple pole expansion in $h_2$ and in $h_3$, with regular parts $U_2$ and $U_3$ respectively. The $U_i$'s are a priori complicated. Instead, we now pass to the simple pole expansion in $c$, which is readily read off from the polar terms in $h_1, h_2, h_3$ (this is very similar to the (\ref{sphtri}) for the sphere 6-point trifundamental block). It then remains to determine the regular part of the conformal block in $c$, which is equivalent to knowing the large $c$ limit.

As we showed in Section \ref{finitec}, the Virasoro conformal block in the plumbing frame built out of of 3-point functions of descendants contracted with inverse Gram matrices has a very simple large $c$ limit: it reduces to the product of the $c\to \infty$ limit of the vacuum Virasoro block and the global $SL(2)$ block (both defined in the plumbing frame). That is,
\ie\label{conjecture}
&\lim_{c\rightarrow\infty}(\text{genus $g$ Virasoro block in channel $\cC$})
\\
=&\lim_{c\rightarrow\infty}(\text{genus $g$ vacuum block in channel $\cC$})\times(\text{genus $g$ global block in channel $\cC$}).
\fe
A genus two example of this was shown in (\ref{genus2largec}).\footnote{We have also confirmed (\ref{conjecture}) for the genus two block (\ref{genus2}) up to total level 12 in $q_1, q_2, q_3$ by scanning over a set of numerical values of the internal weights $h_i$ with Mathematica.}

As already pointed out, the global $SL(2)$ block is easy to compute explicitly in any channel. It is less obvious how to determine the vacuum Virasoro block in the $c\to \infty$ limit on a genus $g$ Riemann surface (since all external and internal primaries are set to identity, there are no more punctures) in a general channel in the plumbing frame, as it receives contributions from all 3-point functions of descendants of the vacuum Verma module. The answer, in fact, is already known, as the holomorphic part of the 1-loop partition function of 3D pure gravity on the corresponding genus $g$ hyperbolic handlebody \cite{Giombi:2008vd}. 

Firstly, note that the vacuum block has the special property that it depends only on the choice of a genus $g$ handlebody that ``fills in" the Riemann surface, i.e. different channels corresponding to the same handlebody (related by crossing moves at the level of sphere 4-point functions) lead to the same answer. In the Schottky parameterization of the moduli, the Riemann surface is realized as a quotient of the form
\ie
(\mathbb{C}\cup \{\infty\} - \Lambda)/\langle \A_1, \cdots, \A_g\rangle,
\fe
where $\A_i$'s are loxodromic elements of $PSL(2,\mathbb{C})$ that act on the Riemann sphere $\mathbb{C}\cup \{\infty\}$ via M\"obius transformation, and $\langle \A_1, \cdots, \A_g\rangle$ is the free group generated by $\A_1,\cdots,\A_g$, known as the Schottky group. $\Lambda$ is the limit set of the Schottky group action. Now given any element $\C$ of the Schottky group, as an element of $PSL(2,\mathbb{C})$ it is conjugate to $\begin{pmatrix} q_\C^{1/ 2} & 0 \\0 & q_\C^{-{1/ 2}} \end{pmatrix}$, with $|q_\C|<1$. Obviously, $q_\C$ depends only on the conjugacy class of $\C$.

Now the $c=\infty$ vacuum block in the pluming frame is given by the product formula \cite{Giombi:2008vd}
\ie
\prod_{\C\in {\cal P}}\prod_{n=2}^\infty(1-q_\C^n)^{-{1\over 2}},
\fe
where ${\cal P}$ is the set of primitive conjugacy classes of the Schottky group. The relation between the Schottky parameters and the plumbing parameters is straightforward. Let us consider as a simple example the genus two partition function. We construct a Virasoro block for the genus two partition function in the plumbing frame by gluing together two 2-holed discs. Gluing one disc into a hole of the other disc leads to a 3-holed disc, where a pair of inner holes are glued together and the remaining inner hole is glued to the outer boundary. The two $PSL(2,\bC)$ maps used in sewing up the boundaries of the 3-holed disc are precisely generators of the Schottky group. The generalization of this procedure to higher genus (involving the gluing of $2(g-1)$ 2-holed discs) is entirely straightforward. Thus, the result of \cite{Giombi:2008vd} combined with the global $SL(2)$ block provide the required $c$-regular part in the plumbing frame, thereby completing the $c$-recursive representation of a general genus $g$ conformal block.

Note that if we move to a different conformal frame, the vacuum block would pick up a conformal anomaly factor, of the form $\exp(-c {\cal F}_0)$, where ${\cal F}_0$ is a function of the moduli. From the holographic perspective, ${\cal F}_0$ is the holomorphic part of the regularized Einstein-Hilbert action evaluated on a genus $g$ hyperbolic handlebody \cite{Krasnov:2000zq, Yin:2007gv}, and the choice of conformal frame is tied to a choice of the cutoff surface along the conformal boundary. The logarithm of the vacuum conformal block is expected to have a $1/c$ asymptotic expansion of the form $-\sum_{k=0}^\infty c^{1-k} {\cal F}_k$, where ${\cal F}_k$ is the holomorphic part of the $k$-loop free energy of the 3D pure gravity at the handlebody saddle point of the gravitational path integral \cite{Giombi:2008vd}. For our purposes here, $\exp(-{\cal F}_1)$ is what survives in the large $c$ limit in the plumbing frame, and serves as the seed that determines the $c$-recursion relation.

To go from the plumbing parameters $q_i$ or the Schottky parameterization of the moduli to the period matrix of the genus $g$ Riemann surface is rather nontrivial (see \cite{Mason:2006dk} for the construction of such a mapping in the genus two case). This is now the main technical obstacle before our recursive representation can be applied to, say, higher genus modular bootstrap.

\section{Discussion}\label{discussion}

In the first part of this work, we derived the $h$-recursion representation of Virasoro conformal blocks for the sphere linear channel and torus necklace channel. The key to this derivation was the determination of the $h$-regular part by taking a simultaneous large weight limit, such that every 3-point function of descendants that appears in the conformal block involves a finite weight primary and a pair of large weights (either primary or descendants), which leads to a drastic simplification of the Virasoro block. Such a limit is not available however for more general conformal blocks, such as the sphere 6-point block in the trifundamental channel.\footnote{In this case, sending 3 of the external weights together with 3 internal weights to infinity while holding their differences fixed indeed still gives a finite limit, but we have not been able to find a closed form expression for the result.} For practical computations, while our $h$-recursive representation does compute order-by-order the expansion of Virasoro block in the plumbing parameters, it is not quite as efficient as that of \cite{1987TMP....73.1088Z}: even in the sphere 4-point case, the residues of the recursive formula involve shifted blocks with a pair of new external weights that now depend on the original internal weight.

The $c$-recursion representations appear to be much more powerful, both in that they apply to arbitrary channel Virasoro conformal blocks on any Riemann surface (provided that we work in the plumbing frame), and they are more efficient for practical evaluation of the $q_i$-expansion.\footnote{For instance, using Mathematica on a laptop, symbolic evaluation of the torus 2-point function in the necklace channel for both $q_1$ and $q_2$ up to level 10 takes typically $\sim\mathcal{O}(10)$ minutes using $c$-recursion, while the same evaluation using $h$-recursion at level 7 takes $\sim\mathcal{O}(10)$ minutes.} It is now possible to compute efficiently the torus 2-point Virasoro blocks in both the necklace channel and the OPE channel, making it possible to analyze the torus 2-point conformal bootstrap for unitary CFTs with semidefinite programming. Note that unlike the conformal bootstrap where crossing symmetry of the sphere four-point function is imposed, here (and generically in higher genus bootstrap) there are multiple internal weights over which the positivity properties must be imposed. This is currently under investigation.

Even though a complete set of consistency constraints on a 2D CFT is captured by the crossing relation of the sphere 4-point function and the modular covariance of the torus 1-point function, the numerical approach to bootstrapping unitary CFTs can hardly incorporate more than a few external operators simultaneously. For this reason it has been rather difficult to combine modular bootstrap and the sphere crossing equation in a useful way. The higher genus conformal bootstrap based on the modular crossing equation would effectively take into account the OPEs of all primaries in the spectrum, without having to work with them individually. This could be very useful especially for theories with large degeneracy/density in the operators.

The remaining complication in implementing higher genus modular bootstrap is to efficiently go between the plumbing parameterization of the moduli and the period matrix, since the latter has a simple modular transformation property while the former transforms in a complicated manner under the modular group. These have been studied in the genus two case in \cite{Mason:2006dk, Yin:2007gv}, but a more efficient computational algorithm will be needed for applications to bootstrap.

Finally, let us mention that our recursive formula allows for the evaluation of torus (and potentially higher genus) correlation functions in Liouville CFT and the $SL(2)$ WZW model, based on integrating a continuous family of conformal blocks with known structure constants. This makes it possible to perform direct numerical evaluation of string loop amplitudes in $c=1$ string theory \cite{Klebanov:1991qa, Ginsparg:1993is}, doubled scaled little string theory  \cite{Giveon:1999px, Aharony:2003vk, Aharony:2004xn, Lin:2015wcg, Lin:2016gcl}, and string theory in $AdS_3$ \cite{Maldacena:2001km}.

\section*{Acknowledgements} 

We would like to thank Chi-Ming Chang, Ying-Hsuan Lin, Alex Maloney, and Eric Perlmutter for discussions. XY thanks Simons Collaboration Workshop on Numerical Bootstrap at Princeton University and ``Quantum Gravity and the Bootstrap" conference at Johns Hopkins University for their hospitality during the course of this work. This work is supported by a Simons Investigator Award from the Simons Foundation and by DOE grant DE-FG02-91ER40654. MC is supported by Samsung Scholarship. SC is supported in part by the Natural Sciences and Engineering Research Council of Canada via a PGS D fellowship. 

\appendix

\section{Virasoro Ward identities}\label{app:review}
The Virasoro algebra is defined by
\ie
\left[L_n, L_m\right]= (n-m)L_{n+m}+{c\over12} n(n^2-1)\D_{n,-m}.
\fe
Ward identities for 3-point functions of Virasoro descendants (\ref{3ptnotation}) were used extensively in \cite{Teschner:2001rv,Hadasz:2006qb, SuchanekThesis}. Here we summarize the relevant results in our notation.
The $z$-dependence of the 3-point function $\rho$ takes the form
\ie
\R(\x_3,\x_2,\x_1|z)=z^{L_0(\x_3)-L_0(\x_2)-L_0(\x_1)}\R(\x_3,\x_2,\x_1|1),
\fe
where $L_0(\x_i)$ is the holomorphic conformal weight of the descendant $\x_i$. Recall that $\xi_1$ is inserted at $z=0$ and $\xi_3$ at $z=\infty$. We have the following Ward identities:
\ie\label{eq:WardIdentities}
\R(\x_3,L_{-1}\x_2,\x_1|z)=&\partial_z\R(\x_3,\x_2,\x_1|z),
\\
\R(\x_3,L_n\x_2,\x_1|z)=&\sum_{m=0}^{n+1} {\begin{pmatrix} n+1 \\ m \end{pmatrix}}
(-z)^m(\R(L_{m-n}\x_3,\x_2,\x_1|z)-\R(\x_3,\x_2,L_{n-m}\x_1|z)),
\\
&~~~~~~~~~~~~~~~~~~~~~~~~~~~~~~~~~~~~~~~~~~~~~~~~~~~~~~~~~~~~~~~~~~~~~~~~n>-1,
\\
\R(\x_3,L_{-n}\x_2,\x_1|z)=&\sum_{m=0}^\infty{\begin{pmatrix}n-2+m \\ n-2\end{pmatrix}}
\left[z^m\R(L_{n+m}\x_3,\x_2,\x_1|z)+(-)^nz^{-n+1-m}\R(\x_3,\x_2,L_{m-1}\x_1|z)\right],
\\
&~~~~~~~~~~~~~~~~~~~~~~~~~~~~~~~~~~~~~~~~~~~~~~~~~~~~~~~~~~~~~~~~~~~~~~~~~n>1,
\\
\R(L_{-n}\x_3,\x_2,\x_1|z)=&\R(\x_3,\x_2,L_n\x_1|z)+\sum_{m=-1}^{l(n)}{\begin{pmatrix}n+1\\m+1\end{pmatrix}}
z^{n-m}\R(\x_3,L_m\x_2,\x_1|z).
\fe
In the last line, $l(n)=n$ for $n\geq-1$, and $l(n)=\infty$ otherwise. In particular, to move $L_m$ acting on $\xi_1$ through a primary $\nu_2(z)$ of weight $d_2$, we can use the commutator
\ie\label{commm}
\left[L_m,\n_2(z) \right]=z^m(z\partial_z+(m+1)d_2)\n_2(z).
\fe

\section{Factorization of 3-point functions}\label{app:3pt}
Here we explain the factorization properties of generic 3-point functions of three descendants involving null states, as given in (\ref{Factorization}).
Using the Ward identities (\ref{eq:WardIdentities}), we can move the Virasoro generators $L_{-M}$ in the second entry of $\rho$ on the LHS of (\ref{Factorization}) to the first and third entries. Thus it suffices to consider the case where $L_{-M}$ is the empty chain. To give a flavor of the derivations, we shall prove (\ref{factorization2}) which is one of the identities in (\ref{Factorization}) with $M=\emptyset$.
Suppose $L_{-A}$ in (\ref{factorization2}) is a Virasoro chain of length $m$, i.e. $L_{-A}=L_{-a_m}L_{-a_{m-1}}...L_{-a_1}$.
The $m=0$ case is easy to prove and was given in \cite{SuchanekThesis}. We will induct on $m$: suppose the property holds for $[A]=m$, and now consider the $[A]=m+1$ case. Repeatedly applying the commutation relation (\ref{commm}), we have
\ie
&\R(L_{-B}\X_{rs},\n_2,L_{-a_{m+1}}L_{-a_{m}}...L_{-a_1}\n_1)
=\R(L_{a_{m+1}}L_{-B}\X_{rs},\n_2,L_{-a_{m}}...L_{-a_1}\n_1)
\\
&~~~~-z^{-a_{m+1}}\left((-a_{m+1}+1)d_2+z\partial_z\right)\R(L_{-B}\X_{rs},\n_2,L_{-a_m}...L_{-a_1}\n_1)
\\
&=\R(L_{a_{m+1}}L_{-B}\n_{d_{rs}+rs},\n_2,L_{-a_{m}}...L_{-a_1}\n_1)\R(\X_{rs},\n_2,\n_1)
\\
&~~~~-z^{-a_{m+1}}\left((-a_{m+1}+1)d_2+z\partial_z\right)\R(L_{-B}\X_{rs},\n_2,L_{-a_m}...L_{-a_1}\n_1)
\\
&=\R(L_{-B}\n_{d_{rs}+rs},\n_2,L_{-a_{m+1}}L_{-a_{m}}...L_{-a_1}\n_1)\R(\X_{rs},\n_2,\n_1)
\\
&~~~~+z^{-a_{m+1}}\left((-a_{m+1}+1)d_2+z\partial_z\right)\R(L_{-B}\n_{d_{rs}+rs},\n_2,L_{-a_m}...L_{-a_1}\n_1)\R(\X_{rs},\n_2,\n_1)
\\
&~~~~-z^{-a_{m+1}}\left((-a_{m+1}+1)d_2+z\partial_z\right)\R(L_{-B}\X_{rs},\n_2,L_{-a_m}...L_{-a_1}\n_1)
\\
&=\R(L_{-B}\n_{d_{rs}+rs},\n_2,L_{-a_{m+1}}L_{-a_{m}}...L_{-a_1}\n_1)\R(\X_{rs},\n_2,\n_1).
\fe
The other identities in (\ref{Factorization}) can be proven similarly, by repeatedly applying (\ref{commm}) and using the property that the null state $\X_{rs}$ behaves as a primary.

Note importantly that the second factors on the RHS of (\ref{Factorization}) are fusion polynomials,
\ie\label{app:fusionpolynomials}
& P^{rs}\begin{bmatrix}d_2\\d_1\end{bmatrix}=\R(\X_{rs},\n_1,\n_2|1)=\R(\n_1,\X_{rs},\n_2|1)=\R(\n_2,\n_1,\X_{rs}|1),
\\
& \R(\X_{rs},\n_1,\X_{rs})=P^{rs}\begin{bmatrix}d_1\\d_{rs}+rs\end{bmatrix}P^{rs}\begin{bmatrix}d_1\\d_{rs}\end{bmatrix}.
\fe

\bibliographystyle{JHEP}
\bibliography{torusdraft}

\end{document}